\documentclass[aps,pra,twocolumn,amsfonts,amssymb,amsmath,10pt,showpacs,showkeys,superscriptaddress]{revtex4-1}
\usepackage{graphicx}
\usepackage{dcolumn}%
\usepackage{bm}%
\usepackage[usenames]{color}
\usepackage{soul} 
\usepackage[normalem]{ulem} 

\begin{document}
\title{Relaxation dynamics of an isolated large-spin Fermi gas far from equilibrium}
\date{\today}

 \author{Ulrich~Ebling}
  \email[Electronic address:]{ulrich.ebling@icfo.es} 
 \affiliation{ICFO - Institut de Ci\`encies Fot\`oniques, Av. Carl Friedrich Gauss, 3, E-08860 Castelldefels (Barcelona), Spain} 
 
 \author{$\!\!{}^{,\,\dagger}\,\,\,$Jasper Simon Krauser}
 \thanks{These two authors contributed equally to this paper}
 \affiliation{Institut f\"ur Laserphysik, Universit\"at Hamburg, Luruper Chaussee 149, D-22761 Hamburg, Germany}

 \author{Nick~Fl\"aschner}
 \affiliation{Institut f\"ur Laserphysik, Universit\"at Hamburg, Luruper Chaussee 149, D-22761 Hamburg, Germany}
 
 \author{Klaus~Sengstock}
 \affiliation{Institut f\"ur Laserphysik, Universit\"at Hamburg, Luruper Chaussee 149, D-22761 Hamburg, Germany}
 \affiliation{ZOQ - Zentrum f\"ur optische Quantentechnologien, Universit\"at Hamburg, Luruper Chaussee 149, 22761 Hamburg, Germany}

 \author{Christoph~Becker}
 \affiliation{Institut f\"ur Laserphysik, Universit\"at Hamburg, Luruper Chaussee 149, D-22761 Hamburg, Germany}
 \affiliation{ZOQ - Zentrum f\"ur optische Quantentechnologien, Universit\"at Hamburg, Luruper Chaussee 149, 22761 Hamburg, Germany}

 \author{Maciej~Lewenstein}
 \affiliation{ICFO - Institut de Ci\`encies Fot\`oniques, Av. Carl Friedrich Gauss, 3, E-08860 Castelldefels (Barcelona), Spain}
 \affiliation{ICREA-Instituci\'o Catalana de Recerca i Estudis Avan\c cats, Llu\'is Companys 23, E-08010 Barcelona, Spain}

 \author{Andr\'e~Eckardt}
 \affiliation{Max-Planck-Institut f\"{u}r Physik komplexer Systeme, N\"othnitzer Str. 38, D-01187 Dresden, Germany}

\begin{abstract}
A fundamental question in many-body physics is how closed quantum systems reach equilibrium. We address this question experimentally and theoretically in an ultracold large-spin Fermi gas where we find a complex interplay between internal and motional degrees of freedom. The fermions are initially prepared far from equilibrium with only a few spin states occupied. The subsequent dynamics leading to redistribution among all spin states is observed experimentally and simulated theoretically using a kinetic Boltzmann equation with full spin coherence. The latter is derived microscopically and provides good agreement with experimental data without any free parameters. We identify several collisional processes, which occur on different time scales. By varying density and magnetic field, we control the relaxation dynamics and are able to continuously tune the character of a subset of spin states from an open to a closed system.
\end{abstract}

\maketitle
\section{Introduction}
\label{sec1}
The relaxation of closed quantum systems towards equilibrium is a fundamental problem in many-body physics. It is particularly challenging to fully understand this macroscopic process on the basis of microscopic properties \cite{Srednicki1994,Rigol2008,Dziarmaga2010,Polkovnikov2011}. Here, ultracold atomic quantum gases provide an exceptional experimental platform due to the nearly perfect isolation from their environment and the excellent control on a microscopic level. In particular, the possibility to prepare well-defined states far from equilibrium, as well as widely tunable Hamiltonians, has recently attracted a lot of attention, e.g. prethermalization \cite{Berges2004,Gring2012,Langen2013}, relaxation in strongly interacting lattice systems \cite{Kinoshita2006,Hofferberth2007,Cheneau2012,Trotzky2012,Lux2013} and the interplay between thermal and condensate fraction of multicomponent bosons have been studied \cite{Erhard2004}.

Spinor quantum gases are of particular interest since the spin offers an additional degree of freedom, giving rise to complex dynamics involving different relaxation processes on different time scales. Ultracold bosonic quantum gases have been intensively studied and exhibit a rich variety of effects such as texture formation and spin dynamics in spinor Bose-Einstein condensates \cite{Schmaljohann2004,Griesmaier2005,Sadler2006,Klempt2009,Kronjaeger2010}, which can be well described theoretically using a multi-component Gross-Pitaevskii-Equation \cite{MurPetit2006,Santos2006,StamperKurn2012}. Recently, collective spin dynamics was also observed in a thermal Bose gas \cite{Pechkis2013}. Fermions, in contrast, are governed by Pauli blocking and reveal a different behavior. So far, most experiments studied spin 1/2 fermions e.g. the BEC-BCS crossover \cite{Regal2004,Zwierlein2005}, thermodynamic and transport properties \cite{Sommer2011,Koschorrek2013}, collective excitations \cite{Du2008,Natu2009} and magnetic ordering \cite{Jo2009,Conduit2011,Pekker2011,Zhang2011}. Spin-related phenomena in multi-component  Fermi gases ($F>1/2$) have been investigated only recently, in the context of spin-mixing dynamics on individual sites of an optical lattice \cite{Krauser2012}, or collective coherent excitations in a trapped system \cite{Heinze2013,Dong2013,Krauser2014}. The latter has been proven to be well described within a Boltzmann equation. The question, how a large-spin fermionic many-body system reaches an equilibrium state via relaxation involving spin and spatial degree of freedom, has not been addressed.

\begin{figure*}[t]
\centering
\includegraphics[width=17.2cm]{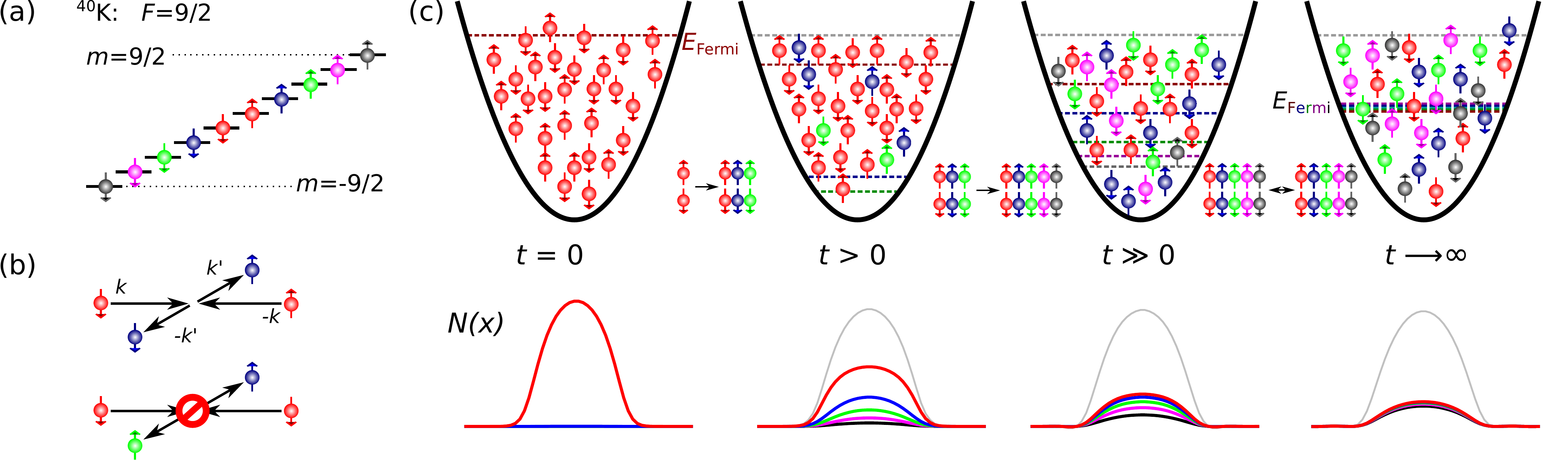}
\caption{Schematic description of the relaxation process in a large-spin Fermi gas involving spin and spatial degrees of freedom. (a) The ten spin states of \textsuperscript{40}K. (b) A typical spin-changing collision in the center-of-mass frame and another collision forbidden by the Pauli exclusion principle. (c) Top: Initially all atoms are prepared in a binary spin mixture $m=\pm1/2$. Spin-changing collisions distribute atoms among all other spin states until an approximately balanced population is reached. The Fermi energies for each two-component subsystem is lower than the initial Fermi energy. Bottom: Time evolution of the spatial density for each spin component.}
\label{fig1}
\end{figure*}
In this paper, we study the relaxation dynamics of a trapped fermionic quantum gas of $^{40}$K atoms with large spin of $F=9/2$. Starting from an initial mixture with only a few spin states occupied, we observe a rich relaxation dynamics leading to a redistribution of the atoms among all available spin states. We study the intermediate regime between the collisionless and the hydrodynamic limit.
In the collisionless limit, interactions are weak and can be taken into account on a mean-field level, while the hydrodynamic limit is characterized by stronger interactions which ensure local equilibrium. The dynamics in this intermediate regime is governed by different processes on very different time scales. We identify these processes by deriving a Boltzmann equation from the microscopic Hamiltonian of the large-spin system. This approach describes the time evolution of the system on the level of single particles in contact with the bath of the many-body system \cite{Gallis1990,Hornberger2003}. This corresponds to the intuitive expectation that the system acts as a bath for its own subsystems.

We present a detailed comparison between numerical simulations and experimental data and find good agreement. Our analysis includes the dependence of the relaxation on density as well as on magnetic field. Whereas a higher density enhances the spin relaxation, we find a suppression of spin-changing processes at large magnetic fields due to the quadratic Zeeman shift. The latter effect can be used to control the loss of particles from the subsystem defined by the initially occupied spin components into the initially empty spin states.
 Generally, we observe that the relaxation within a subset of spin states, driven by incoherent spin-conserving collisions, happens on a much faster time scale than the redistribution among the spin components due to spin-changing incoherent collisions. The reason is that the spin-changing collisions are driven by the relatively small part of the interactions that breaks the $SU(10)$ symmetry between the spin states. Thus, we encounter a situation similar to prethermalization \cite{Berges2004}, where first a prethermal state is reached, approximately conserving the initial occupations of the single spin states, before the redistribution among all spin states due to slight symmetry breaking sets in. This separation of time scales also allows us to monitor the increase of (effective) temperature within the subsystem of the initially populated spin states, as it is caused by dissipation into empty spin states.

\section{Relaxation processes in a large-spin system} 
\label{sec2}
We perform measurements in a quantum degenerate gas of fermionic \textsuperscript{40}K, which has total spin $F=9/2$ in its hyperfine ground state, yielding ten spin states $m=-9/2\ldots+9/2$, as depicted in FIG.~\ref{fig1}(a). We prepare an atomic sample with two spin states occupied (see Appendix \ref{app:A1} for details), confined in a spin-independent dipole trap. Due to the broken $SU(N)$ symmetry in \textsuperscript{40}K resulting from spin-dependent scattering lengths, spin-changing collisions can occur. A microscopic collision process is depicted in FIG.~\ref{fig1}(b): Two particles collide and exchange both spin $m$ and momentum $k$: $(|m_1,k_1\rangle+|m_2,k_2\rangle \rightarrow |m'_1,k'_1\rangle+|m'_2,k'_2\rangle)$. The total spin $S$, the total magnetization $M=m_1+m_2$ as well as the total momentum $k_1+k_2$ have to be conserved in this process. As a particular fermionic feature, the Pauli exclusion principle has to be obeyed, i.e. $m_1 \neq m_2$ and $m'_1 \neq m'_2$. The interplay between the differential quadratic Zeeman energy $\propto m_1^2+m_2^2-{m'}^2_1-{m'}^2_2$ and interaction energy determines whether spin-changing collisions are likely or suppressed. In the presence of spin-changing collisions, the atoms will in general relax into a steady state with population in all ten spin states. Hence, preparation of an initial non-equilibrium state with only a few spin states populated will lead to complex dynamics, in which more and more spin states are gradually occupied (see FIG.~\ref{fig1}(c)). It is a compelling question how the system relaxes towards a steady state.

\begin{figure}[t]
\centering
\includegraphics[width=8.6cm]{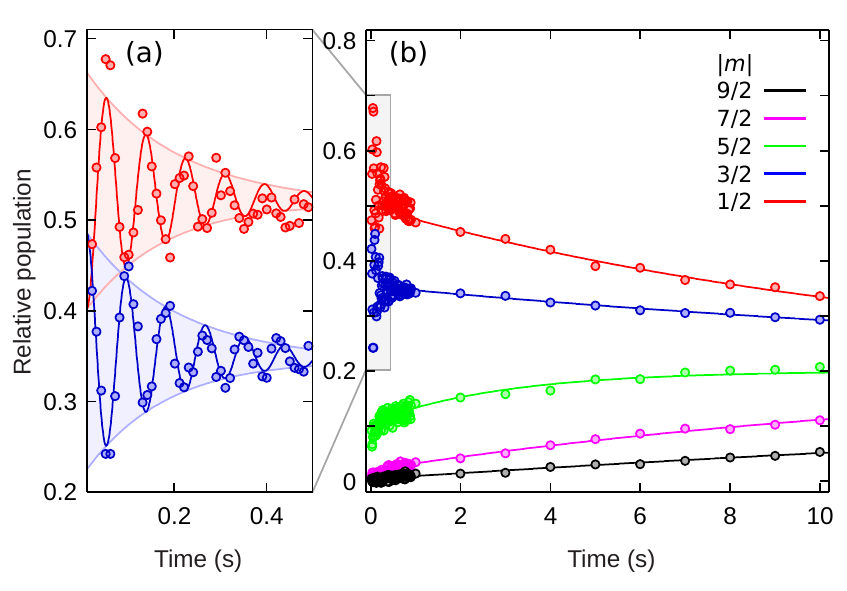}
\caption{(a) Measurement of damped spin oscillations and subsequent relaxation towards equilibrium (b), observed in a 3D fermionic quantum gas with large spin. Depicted is the time evolution of the relative populations of all spin-components $\pm m$, starting from an initial superposition of all ten spin states. For the exact experimental configuration, see Appendix \ref{app:A1}. Solid lines are guides-to-the-eye. Note the three time scales of (i) the spin oscillations, (ii) their damping and (iii) the subsequent relaxation of the total system. The redistribution among all spin states occurs on a time scale of $10\,\text{s}$. The magnetic field is $B=0.17\,\text{G}$, particle number $N=4.9\times10^5$ and temperature $T/T_\text{F}=0.22$.}
\label{fig2}
\end{figure}

In FIG.~\ref{fig2} we show exemplarily an experimentally obtained time evolution of the spin occupations in our system. Here, the initial spin configuration is a superposition of all ten spin states created by rotating a mixture of the states $m=\pm1/2$ using rf-pulses \cite{Krauser2014}.
We can clearly identify three different processes occurring on three different time scales: 
(i) We observe coherent spin-changing oscillations with a periodicity on the order of hundred ms. (ii) These oscillations are damped with a rate on the order of several hundred ms. (iii) Beyond this, we observe a slow redistribution among the ten spin states on a much longer time scale on the order of tens of seconds. 

In the following, we derive a Boltzmann equation, which reproduces the experimentally observed effects and enables us to distinguish, which scattering processes are responsible for each effect. We show that the coherent oscillation (i) are a mean-field effect driven by forward scattering, where $\lbrace k_1,k_2\rbrace=\lbrace k'_1,k'_2\rbrace$. Their damping (ii) is dominated by spin-conserving non-forward collisions $\lbrace k_1,k_2\rbrace\neq\lbrace k'_1,k'_2\rbrace$ and $\lbrace m_1,m_2\rbrace=\lbrace m'_1,m'_2\rbrace$, which lead to a momentum redistribution within the Fermi sea without changing the spin configuration. The long-term redistribution (iii) is governed by non-forward spin-changing collisions, which change the momentum distribution within the Fermi sea as well as the spin configuration $\lbrace k_1,k_2\rbrace\neq\lbrace k'_1,k'_2\rbrace$ and $\lbrace m_1,m_2\rbrace\neq\lbrace m'_1,m'_2\rbrace$. 
\begin{figure}[t]
\centering
\includegraphics[width=8.6cm]{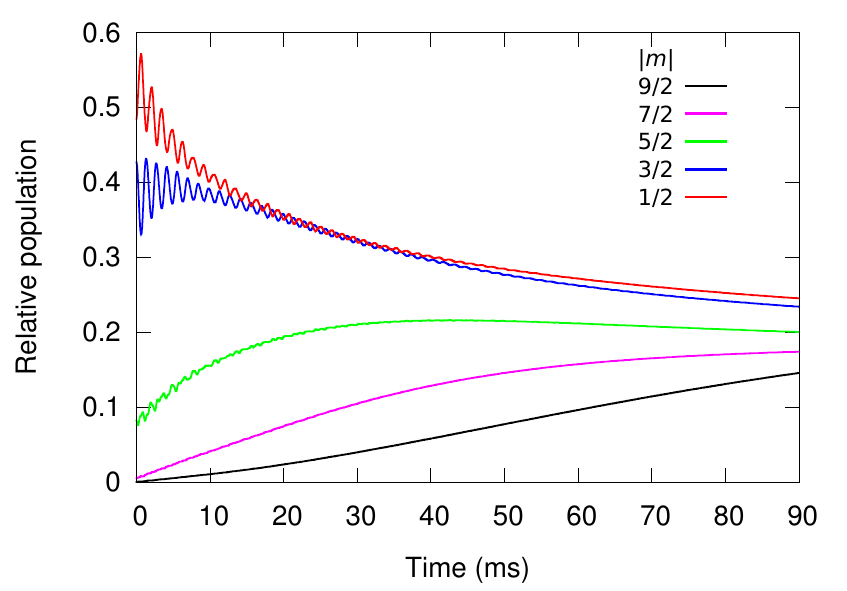}
\caption{Numerical simulation of coherent oscillations, damping and relaxation in the 1D case. The initial spin configuration is the same as in FIG.~\ref{fig2}. Axial trapping frequency is $\omega_x=2\pi\times 84\,\text{Hz}$ and radial frequencies are $\omega_{y,z}=2\pi\times 47\,\text{kHz}$, particle number $N=100$ per tube at temperature $T/T_\text{F}=0.2$ and magnetic field $B=1.5\,\text{G}$. As in FIG.~\ref{fig2}, we observe three time scales related to oscillations, damping and relaxation.}
\label{fig3}
\end{figure}

The abovementioned Boltzmann equation includes all these collision processes and captures the non-equilibrium dynamics in a general fashion, applicable to trapped weakly-interacting gases with arbitrary spin. In this approach, the single-particle dynamics is treated as an open system in contact with the environment represented by all the other particles. An approach for deriving a Boltzmann equation was applied successfully to the description of spin dynamics in liquid Hydrogen and Helium \cite{Corruccini1972,Johnson1984,Bashkin1981,Levy1984,Lhuillier1982I,Lhuillier1982II,OwersBradley1997} and later in spin 1/2 Fermi gases \cite{Fuchs2003, Piechon2009}. In this paper we generalize this approach to one- and three-dimensional systems with large spin, accounting for the quadratic Zeeman effect (QZE).
In general, a kinetic equation or Boltzmann equation is used to describe the time evolution of the single-particle density matrix $\hat\rho$. It has the form
 \begin{equation}
  \frac{d}{dt}\hat\rho-\frac{1}{i\hbar}\left[\hat\rho,\hat H_0\right]=I_\text{coll}\left[\hat\rho\right].
  \label{eq:1}
 \end{equation}
Here, $\hat H_0$ denotes the single-particle Hamiltonian
\begin{equation}
 \hat H_0=\frac{\hat p^2}{2M}+\frac12 M\omega^2\hat x^2+Q\hat S_z^2,
 \label{eq:h0}
\end{equation}
which contains the kinetic energy, the harmonic trapping potential and the quadratic Zeeman splitting $Q$ induced by a homogeneous magnetic field. The term $I_\text{coll}\left[\hat\rho\right]$ on the right hand side of (\ref{eq:1}) is called the collision term and is derived from two-particle contact interaction. Due to total spin conservation, collisions are best described in the basis of total spin $|S,M\rangle=\sum_{m_1,m_2} |m_1,m_2\rangle\langle m_1 m_2|SM\rangle$ with the short notation for Clebsch-Gordan coefficients $\langle m_1 m_2|SM\rangle\equiv\langle F,m_1;F,m_2|S,M\rangle$, which we use throughout this paper. In general, scattering in each channel of total spin $S$ depends on a different s-wave scattering length $a_S$. Due to antisymmetrization of the total wave function, s-wave scattering with odd $S$ is forbidden. Thus for ${}^{40}\mathrm{K}$ there are five different scattering lengths present for $S=0,2,4,6,8$ \cite{Krauser2012}. In each collision channel defined by $|S,M\rangle$, particles interact with a contact interaction of strength $g_S^\text{3D}=\frac{4\pi\hbar^2}{M}a_S$, which is used in all 3D calculations. We also consider a 1D system, in which the motion in two transversal directions is frozen out completely by a tight trapping potential (characterized by radial frequencies $\omega_{x,y}$) such that the effective 1D contact interaction parameter is given by $g_S^\text{1D}=2\hbar\sqrt{\omega_y\omega_z}a_S$ (see appendix~ \ref{app:A1} for details). The notation $g_S\equiv g^\text{1D}_S$ for this quantity is used throughout this paper. In \textsuperscript{40}K, the $a_S$ range from 120 to 170 Bohr radii.

We obtain an explicit expression for the collision term in (\ref{eq:1}) using the method originally developed by Lhuillier and Lalo\"e \cite{Lhuillier1982I,Lhuillier1982II,Fuchs2003} for transport properties in Helium. In this approach, collisions are treated as single ``atomic beam'' experiments, where the colliding particles are assumed to be uncorrelated before and after a collision, reminiscent of Boltzmann's original molecular chaos hypothesis, but the scattering process itself is treated on a full quantum level. This approximation is valid for dilute gases where the mean time between collisions is long and the particle number is large. In this regime, binary collisions can be described by the $T$-matrix, which connects the two-body density matrices before and after a collision. Subsequently, the description is reduced to a single-particle level by tracing out the second particle, similar to tracing out a thermal bath in studies of collisional decoherence \cite{Gallis1990,Hornberger2003}.

We calculate the kinetic equation (\ref{eq:1}) in its phase-space representation, where the single-particle density matrix $\rho_{mn}(x,x')$ is expressed by the Wigner function 
\begin{equation}
W_{mn}(x,p)=\int\frac{dy}{2\pi\hbar}e^{\frac{ipy}{\hbar}}\rho_{mn}(x+\tfrac{y}{2},x-\tfrac{y}{2}).
\end{equation}
Note that we performed the transformation only with respect to the spatial degrees of freedom. With respect to spin, denoted by the indices, it retains the form of a single-particle density matrix.
The derivation is carried out in detail in section \ref{sec4} and involves a semiclassical gradient expansion of the Wigner function in position and momentum space leading to an equation in matrix form given by
\begin{widetext}
\begin{equation}
\frac{d}{dt} W(x,p)+\partial_0 W(x,p)+\frac{i}{\hbar}\left[Q S_z^2+V^\text{mf}(x), W(x,p)\right]-\frac12\left\lbrace\partial_x V^\text{mf}(x),\partial_p W(x,p)\right\rbrace=I_\text{coll}(x,p)
\label{eq:4}
\end{equation}
where the collision integral reads
\begin{align} 
I_{mn}^\text{coll}&(x,p)=-\frac{M}{\hbar^2}\sum_{abl}\left\lbrace\int_{q^2>\epsilon_1}\!\!\!dq\frac{\tilde U_{malb}}{\sqrt{q^2+\Delta_{mlab}}}W_{an}(x,p)W_{bl}(x,p-q)+\int_{q^2>\epsilon_2}\!\!\!dq\frac{\tilde U_{nalb}}{\sqrt{q^2+\Delta_{nlab}}}W_{ma}(x,p)W_{lb}(x,p-q)\right\rbrace\nonumber\\*
&+\frac{M}{\hbar^2}\sum_{abcdl}\int_{q^2>\epsilon_3} dq\frac{U_{malb} U_{ncld}}{\sqrt{q^2+\Delta_{mnlabcd}}} W_{ac}(x,p-\tfrac12 (q-\sqrt{q^2+\Delta_{mnlabcd}}))W_{bd}(x,p-\tfrac12(q+\sqrt{q^2+\Delta_{mnlabcd}})).
\label{eq:5}
\end{align}
\end{widetext}
Here, we define the coupling constants 
 \begin{align}
 U_{acbd}=\sum_{S,M}g_S\left\langle ab|SM\right\rangle\left\langle SM|cd\right\rangle\\*
 \tilde U_{acbd}=\sum_{S,M}g^2_S\left\langle ab|SM\right\rangle\left\langle SM|cd\right\rangle
 \end{align}
and denote energy shifts induced by the quadratic Zeeman splitting as  $\Delta_{abcd}=4MQ(a^2+b^2-c^2-d^2)$ and $\Delta_{mnlabcd}=2MQ(m^2+n^2+2l^2-a^2-b^2-c^2-d^2)$. The infrared cutoffs are given by $\epsilon_1=\frac{MU_{malb}}{\hbar}-\Delta_{mlab}$, $\epsilon_2=\frac{MU_{nalb}}{\hbar}-\Delta_{nlab}$ and $\epsilon_3=\frac{MU_{nalb}}{2\hbar}(U_{malb}+U_{ncld})-\Delta_{mnlabcd}$.

Equation (\ref{eq:4}) contains several terms, each describing a different dynamical process. The free particle motion in the trap is described by $\partial_0=\frac pM\partial_x-M\omega_x^2x\partial_p$. The leading interaction term appears in the commutator  $[\cdot,\cdot]$. The commutator drives coherent spin dynamics through the interplay of the quadratic Zeeman effect and a spin-dependent mean-field potential resulting from forward scattering:
\begin{equation}
\label{eq:8}
 V_{mn}^\text{mf}(x)=2\sum_{ab}U_{mnab}N_{ab}(x),
\end{equation}
a function of the density $N(x)=\int dp W(x,p)$. In our large-spin system, described by several scattering lengths $a_S$, it is helpful to decompose the mean-field potential (\ref{eq:8}) into two contributions. The first contribution is symmetric with respect to all $N$ spin states and proportional to the mean scattering length. It conserves the occupations of the different spin components. The second term contains that part of the interactions that breaks the $SU(N)$ symmetry between the spin states and describes spin-changing processes. It depends on differences of scattering lengths only and is, thus, typically much smaller than the symmetric term. 
The commutator in Eq.~(\ref{eq:4}) vanishes unless the Wigner function possesses off-diagonal elements indicating spin coherence. Moreover, the symmetric spin-conserving mean-field interactions can only contribute if the Wigner function describes an inhomogeneous spin state.

The mean-field potential also appears in the anticommutator $\lbrace\cdot,\cdot\rbrace$. This term results from the subleading order of the semiclassical gradient expansion (where also the spin-independent trapping potential appears) and it is generally smaller than the commutator. It describes spin-dependent forces, that modify the kinetics in the trap. The collision integral (\ref{eq:5}) describes effects beyond mean-field that result from non-forward scattering and generates a dynamics that appears incoherent on the level of a single-particle description. It is quadratic in the scattering lengths. Again we have to distinguish between $SU(N)$ symmetric spin-conserving collision processes on the one hand and spin-changing collisions on the other. The latter processes are described by those terms for which the quadratic Zeeman shifts $\Delta$ are non-zero and they are much smaller than the former, since they depend on the relatively small differences between the scattering lengths only.

\begin{figure}[t]
\centering
 \includegraphics[width=8.6cm]{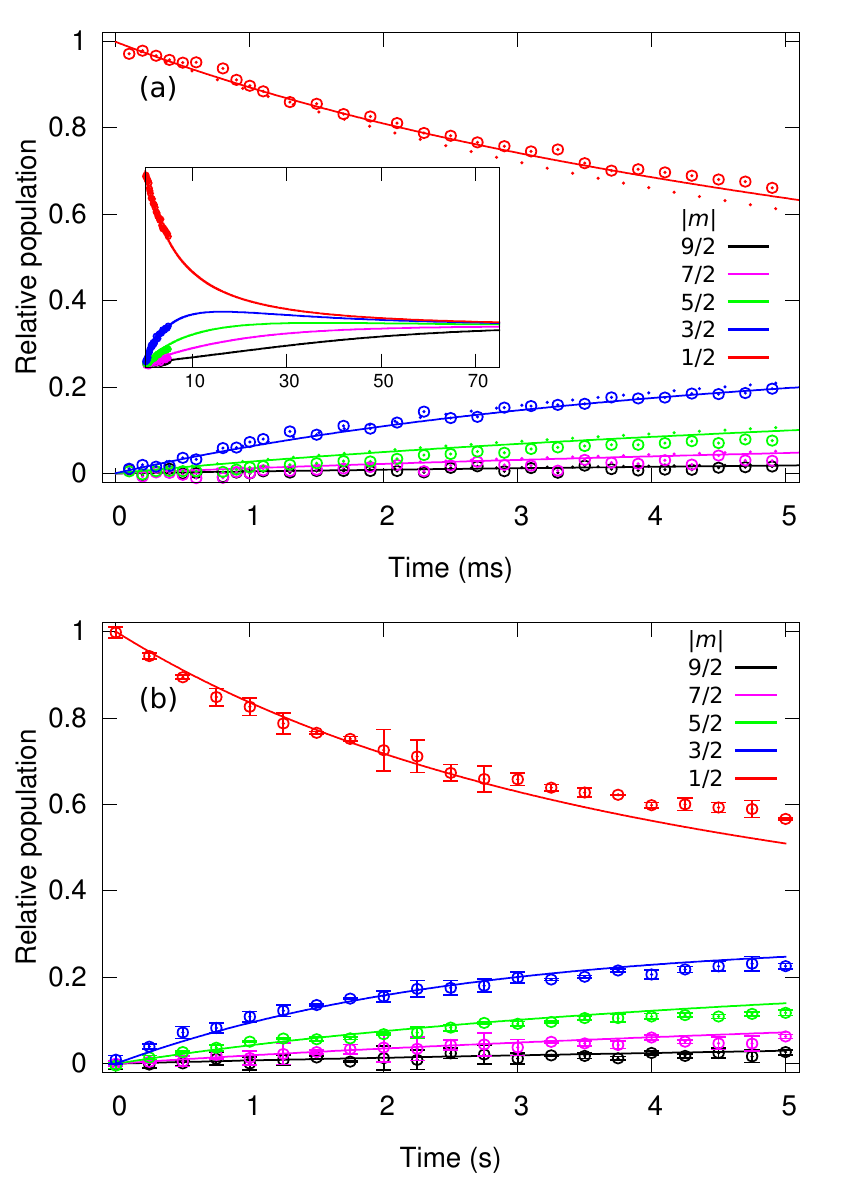}
\caption{Comparison of spin relaxation for 1D and 3D. The initial spin configuration is a mixture of $m=\pm1/2$. (a) Experimental data in a 1D geometry (circles) compared to numerical results (lines) from the 1D Boltzmann equation (\ref{eq:4}) and (dots) from a 1D version of the single-mode approximation (\ref{eq:sma}). Axial trapping frequency is $\omega_x=2\pi\times 84\,\text{Hz}$ and radial frequencies are $\omega_{y,z}=2\pi\times 47\,\text{kHz}$, particle number $N=100$ per tube at temperature $T/T_\text{F}=0.2$ and magnetic field $B=0.12\,\text{G}$. Inset: The system approaches a steady state for longer times. (b) Experimental data (circles) in a 3D configuration compared to calculations (lines) in single-mode approximation (\ref{eq:sma}), $\vec\omega=2\pi\times(33,33,137)\,\text{Hz}$, $N=1.3\times10^5$ and $T/T_\text{F}=0.15$ at $B=0.34\,\text{G}$.}
\label{fig4}
\end{figure}

\begin{figure}[t]
\centering
\includegraphics[width=8.6cm]{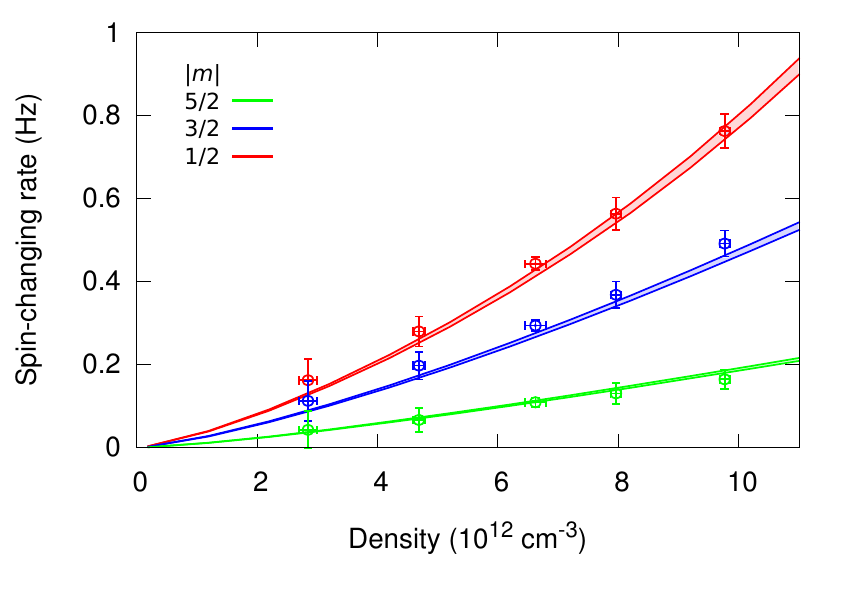}
\caption{Density dependence of the spin relaxation rate in 3D, with an initial mixture of atoms in $m=\pm 1/2$. The spin-changing rate is obtained by fitting the solution of coupled rate equations to experimental (points) and theoretical (lines) data. Theoretical values are obtained using the single-mode approximation using Eq.~(\ref{eq:sma}). The magnetic field is $B=0.11\,\text{G}$. We experimentally tune the density by changing the particle number, keeping the temperature constant at $T/T_\text{F}=0.26$.}
\label{fig5}
\end{figure}

The 1D equation (\ref{eq:4}) allows for a numerical treatment with standard methods. An exemplary result is depicted in FIG.~\ref{fig3}, and shows the relaxation dynamics starting from the same initial superposition as in FIG.~\ref{fig2}, but in a 1D setup. The comparison of the two figures allows us to assign each of the three different processes visible to one of the terms in the Boltzmann equation:
(i) The coherent oscillations are driven by the commutator in (\ref{eq:4}), which is linear in differences of scattering lengths and describes forward collisions. (ii) The damping of coherent phenomena arises from the spin-conserving part of the collision integral (\ref{eq:5}), which is quadratic in the scattering lengths. We have checked that spatial dephasing is not responsible as it is suppressed by the dynamically induced long-range nature of mean-field interactions induced by the rapid particle motion in the trap \cite{Ebling2011,Krauser2014}. (iii) The long-term relaxation originates from the spin-changing non-forward collisions in the collision term, quadratic in differences of scattering lengths. Spin-conserving forward scattering does not play a role in the dynamics, it only has a noticeable effect, if spatial symmetry is broken by a magnetic field gradient, as in studies of spin-waves \cite{Fuchs2003,Natu2009,Heinze2013}, which is not the case in our setup. The anticommutator in Eq.~(\ref{eq:4}) leads to a mean-field driven correction to the trapping potential which is however negligible in the experiments considered here.

The collision integral (\ref{eq:5}) enables us to determine whether our system is in the collisionless, hydrodynamic or an intermediate regime. The average collision time in the 3D setup $\tau_\text{3D}\sim (4\pi a^2 n_p v_T)^{-1}$  \cite{Piechon2009}, with the relevant scattering length $a$, peak density $n_p$ and thermal velocity $v_T=\sqrt{k_BT/M}$, ranges from $\sim 30\,\text{ms}$ to $\sim 150\,\text{ms}$ for spin-conserving collisions and $\sim 3\,\text{s}$ to $\sim15\,\text{s}$ in the spin-changing case. Compared to the average trapping frequency of
$\bar\omega=(\omega_x\omega_y\omega_z)^{1/3}\approx2\pi\times 58\,\text{Hz}$, we obtain values for $\bar\omega\tau_\text{3D}$ between $11$ and $55$ for the spin-conserving collisions and between $1100$ and $5500$ in the spin-changing case. The lowest and highest values of $\bar\omega\tau_\text{3D}$ are reached for the lowest and highest densities shown in FIG.~\ref{fig5} respectively. This means we may approach the hydrodynamic regime, where the collision rate is larger than $\bar\omega$ and local equilibrium can be established. On the other hand our system becomes almost collisionless regarding the spin-changing collisions. Generally we are in an intermediate regime. In the 1D case, collision times $\tau_\text{1D}\sim (n_p\omega_y\omega_z a^2/v_T)^{-1}$ are on the order of $0.35\,\text{ms}$ and $35\,\text{ms}$ respectively, meaning that $\omega\tau_\text{1D}\sim 0.2$ and $\omega\tau_\text{1D}\sim 20$, concerning spin-conserving and spin-changing collisions respectively. Hence with respect to the former, the system would be hydrodynamic. However, it is still in an intermediate regime regarding the redistribution of particles among the spin states driven by spin-changing collisions.

\section{Dissipative redistribution of spin occupations}
\label{sec3}
In the following, we focus on the long-term spin relaxation shown in FIG.~\ref{fig2}(b) and FIG.~\ref{fig3}, while recent experiments have studied spin oscillations and their damping \cite{Krauser2014}. In order to restrict the dynamics to this process, we initially prepare a spin mixture consisting of only the spin states $m=\pm1/2$ without coherences. In this case, the coherent oscillations driven by the commutator in Eq.~(\ref{eq:4}) are absent and the Wigner function $W_{mn}$ remains diagonal at all times. In the following we investigate theoretically and experimentally this spin relaxation dynamics in 3D as well as 1D systems. For a direct comparison between theory and experiment, we realize a 1D system employing a deep 2D optical lattice, which confines the atoms into tight elongated tubes \cite{Fertig2005,Kinoshita2006} as described in Appendix \ref{app:A1}.  As shown in FIG.~\ref{fig4}(a), the system  gradually occupies all spin states and evolves towards a state of almost equal spin populations on a time scale of milliseconds. As a key result, we can well reproduce the experimentally observed dynamics using the full 1D Boltzmann equation without free parameters.  

For harmonically trapped 3D systems, where all trap frequencies are about equal, we derive the full 3D version of equation (\ref{eq:4}) as well [see Eq.~(\ref{eq:28}) in section \ref{sec4}]. However, numerical simulations of this equation are too demanding numerically. Nevertheless, the trap-induced motion of the particles is considerably faster than mean-field or relaxation dynamics, which averages the spatial dependence of the interaction via dynamically created long-range interactions. The Wigner function then approximately separates into a product $W_{mn}(\bm x,\bm p,t)\approx M_{mn}(t)\cdot f_0(\bm x,\bm p)$  \cite{Ebling2011,Pechkis2013,Krauser2014}. The spatial part is assumed to be time independent and given by the initial equilibrium distribution $f_0(\bm x,\bm p)=(\exp(\frac{1}{k_BT}[\frac{\bm p^2}{2M}+\frac12 M (\omega_x^2 x^2+\omega_y^2 y^2+\omega_z^2 z^2)-\mu])+1)^{-1}$. We substitute this expression into the 3D kinetic equation (\ref{eq:28}) with the appropriate collision term. Hence, for negligible magnetic fields we find an equation for the matrix $M_{mn}(t)$ given by 
\begin{equation}
 \frac{d}{dt}M_{mn}=-\lambda\sum_{abcd} T_{mn}^{abcd} M_{ac}M_{bd},
 \label{eq:sma}
\end{equation}
where 
\[T_{mn}^{abcd}=\frac{M}{4\pi\hbar^4}\left(\tilde U'_{mabd}\delta_{nc}+\tilde U'_{ncbd}\delta_{ma}-\sum_l U'_{malb} U'_{ncld}\right)\]
and 
\begin{equation}
 \lambda=\frac1N\int d\bm r\int d\bm p\int d\bm q |\bm q|f_0(\bm r,\bm p)f_0(\bm r,\bm p-\bm q).
 \label{eq:lambda}
\end{equation}
The 3D coupling constants are given by 
\begin{equation}
U'_{acbd}=\frac{4\pi\hbar^2}{M}\sum_{S,M}a_S\left\langle ab|SM\right\rangle\left\langle SM|cd\right\rangle
\end{equation}
and
\begin{equation}
\tilde U'_{acbd}=\frac{16\pi^2\hbar^4}{M^2}\sum_{S,M}a^2_S\left\langle ab|SM\right\rangle\left\langle SM|cd\right\rangle.
\end{equation}
Note that in the 1D case the single-mode approximation has a similar form given by $T_{mn}^{abcd}=\frac{M}{\hbar^2}\left(\tilde U_{mabd}\delta_{nc}+\tilde U_{ncbd}\delta_{ma}-\sum_l U_{malb} U_{ncld}\right)$.

For both single-mode approximations, the quadratic Zeeman shift has been neglected in the above equations, which are thus valid for small magnetic fields only (See Appendix \ref{app:G} for full equations). In FIG.~\ref{fig4} we compare results from a single-mode approximation with experiments in a harmonically trapped Fermi gas yielding a surprisingly good agreement without free parameters. Note the qualitatively comparable behavior on different time scales of milliseconds for 1D and seconds for 3D. On the contrary the damping of the coherent spin oscillations visible in FIG.~\ref{fig2}(a) and FIG.~\ref{fig3} is not described by this approach. This results from the assumption for the single-mode approximation that it completely neglects the multi-mode character of the fermionic many-body system and thus cannot account for spatial redistribution via lateral scattering events. 

The high degree of control over all crucial parameters allows for a detailed investigation of the spin redistribution. To obtain further insight into the relaxation mechanisms, we measure the relaxation dynamics (as exemplarily shown in FIG.~\ref{fig4}) for different densities while keeping $T/T_\text{F}$ constant. With higher density the collision rate increases and the relaxation process accelerates, as shown in FIG.~\ref{fig5}. The measured rates correspond to the redistribution of the initially populated components $m=\pm 1/2$ into $m=\pm 3/2,\pm 5/2$ and are well reproduced using the single-mode approximation (\ref{eq:sma}). The rate of spin-changing collisions increases with increasing density in accordance with the density dependence of the integral $\lambda$ (\ref{eq:lambda}).

\begin{figure}[t]
\includegraphics[width=8.6cm]{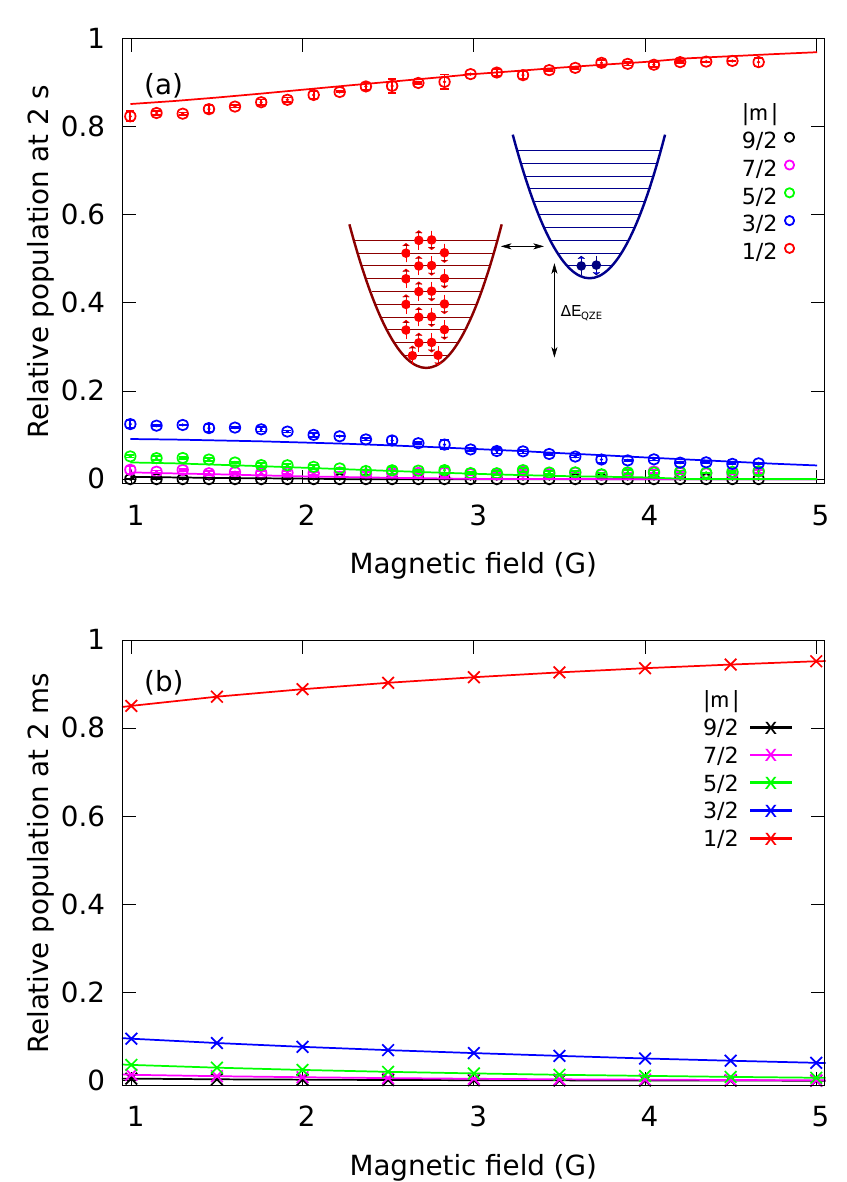}
\caption{Dependence of spin relaxation on magnetic field. (a) Experimental data, obtained from a 3D experiment (circles) and theoretical results from a single-mode approach (lines). Spin populations are measured after $2\,\text{s}$. (b) Spin populations after $2\, \text{ms}$, as obtained from full 1D simulations. The inset sketches how the interplay of differential QZE and Fermi energy determines the probabilities for lateral spin-changing collisions.}
\label{fig6}
\end{figure}

\begin{figure}[t]
\centering
\includegraphics[width=8.6cm]{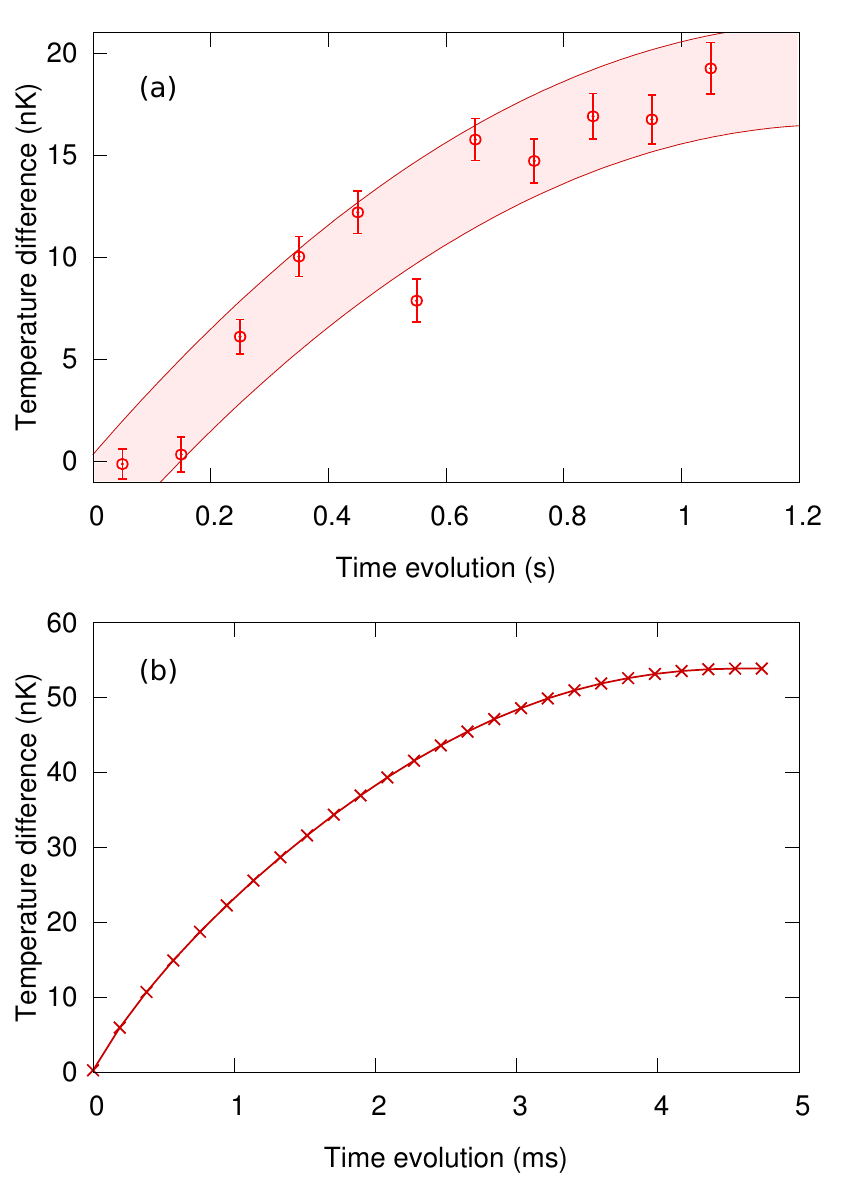}
\caption{Temperature increase due to spin redistribution. (a) Time evolution of the temperature difference between a closed (high magnetic field at $B=7.6\,\text{G}$) and a maximally open system (low magnetic field at $B=0.12\,\text{G}$). The shaded area serves as a guide-to-the-eye. The particle number is $N=3.9\times 10^5$ at an initial temperature $T=0.24\,T_\text{F}=65\,\text{nK}$. (b) Results from a simulation of the 1D equation (\ref{eq:4}) at magnetic fields $B=0.1\,\text{G}$ and $B=8\,\text{G}$. The method used to obtain the temperature is discussed in Appendix \ref{app:H}.}
\label{fig7}
\end{figure}

As a second important parameter of the system, we investigate the influence of the magnetic field on the relaxation process. As the Zeeman energy of an atom pair changes during a spin-changing collision, a strong magnetic field suppresses this process by increasing the energy difference between the initial and final spin configuration. In FIG.~\ref{fig6}(a), we depict the experimentally obtained populations of the spin components after $2\,\text{s}$ as a function of the magnetic field strength (see Appendix \ref{app:A2} for details) and compare them to single-mode [FIG.~\ref{fig6}(a)] and 1D calculations [FIG.~\ref{fig6}(b)] after $2\,\text{ms}$. In both cases, the general behavior is very similar and shows a suppression of spin-changing collisions for large magnetic fields. Spin configurations with high values of $|m|$ are energetically significantly separated from the initially populated $m =\pm{1/2}$ and are only occupied at very low field strengths. By changing the magnetic field we can thus tune the magnitude of spin-changing collisions relative to the unaffected spin-conserving collisions up to a complete suppression. This gives us the possibility to view the $m=\pm1/2$ subsystem as a dissipative two-component Fermi gas with a tunable loss mechanism.

We have further investigated the time evolution of the temperature of this subsystem exposed to losses induced by these  spin- and momentum changing collisions. We compare two experiments: On the one hand, we perform an experiment at a high magnetic field ($B=7.6\,\text{G}$), where spin-changing collisions are suppressed. On the other hand, we perform a second experiment at a low magnetic field ($B=0.12\,\text{G}$) with strong spin relaxation. 

In both cases the system is in the hydrodynamic limit with respect to external degrees of freedom due to the comparatively large spin-conserving interactions. Hence we make the assumption, that at each time the subsystems are close to an intermediate equilibrium state with a well-defined temperature. On the other hand, spin-changing collisions are at least two orders of magnitude weaker and very slowly change the particle number in each subsystem.  This situation is reminiscent of prethermalization, first a "prethermal" state is reached under the assumption of conserved quantities, which on a much longer time scale are actually not fully conserved due to a "slightly broken symmetry", leading eventually to full thermalization \cite{Berges2004}. Here, the role of the nearly conserved quantities is played by the occupation numbers of the ten spin states, which change only on a very long time scale. We measure the time evolution of the temperature of the initially populated $m=\pm1/2$ components and compare the temperatures for both cases described above. For large magnetic fields we observe a small heating rate, which we mainly attribute to inelastic photon scattering. However, at low magnetic fields, the heating rate is significantly increased. In FIG.~\ref{fig7}(a), we plot the temperature difference to extract the heating contributions solely generated by spin-changing collisions. This additional increase in temperature is due to hole creation in the Fermi sea \cite{Timmermans2001} by scattering into the unoccupied spin states. We initially prepare a very cold two-component Fermi sea, with only few unoccupied trap levels below the Fermi energy. Losses through spin-changing collisions ``perforate'' this Fermi sea with holes, such that the experimentally obtained temperature increases.
Numerical simulations using the Boltzmann equation (\ref{eq:4}) confirm the experimentally observed heating induced by redistribution [see FIG.~\ref{fig7}(b)].

\section{Microscopic derivation of a large-spin Boltzmann equation}
\label{sec4}
In this section, we describe the derivation of the 1D Boltzmann equation (\ref{eq:4}) in more detail. The reader not immediately interested in the details may skip this section and proceed directly to the conclusions. We follow previous work on the theoretical description of spin-polarized systems of $\text{H}$ or $\text{He}$ \cite{Lhuillier1982I}, called the Lhuillier-Lalo\"e transport equation. We extend it to describe a 1D system with large spin, several scattering channels and a quadratic Zeeman effect. We consider this approach to be suitable for our purpose for a couple of reasons. The entire equation is derived from a microscopic collisional approach, so the collision term is not based on phenomenological assumptions. We avoid the use of a relaxation time approximation, widely used to describe damping in bosonic and fermionic systems \cite{Kavoulakis1998,Massignan2005,Natu2010}, where the collision term is approximated by $I_\text{coll}\left[W\right]=\frac{W_\text{eq}- W}{\tau}$, with a relaxation time $\tau$. The reason is that for a multicomponent system determining the equilibrium state $W_\text{eq}$ is very challenging. Also due to the interplay of many different scattering lengths we expect not one but many different relaxation times for each spin component. Our approach allows to better understand the relaxation process itself, rather than merely its effect on other processes. Furthermore, from a technical point of view, our approach remains quadratic in the Wigner function so it can be numerically simulated using the same standard techniques as the collisionless case \cite{Piechon2009,Natu2009,Ebling2011}. 

The idea behind the approach of Lhuillier-Lalo\"e is to interpret the collision integral as the change rate of the state of a single particle $\hat\rho\rightarrow\hat\rho'$ due to binary collisions
\begin{equation}
I_\mathrm{coll}=\frac{\hat\rho'-\hat\rho}{\Delta t}.
\end{equation} 
Here $\Delta t$ is the elapsed time interval, which is short compared to any relevant macroscopic dynamics of the system, but nevertheless is longer than the duration of a single collision, which is thus considered to be effectively instantaneous. This quantity will drop out and not appear in the final kinetic equation. With this in mind, we treat collisions in the asymptotic limit, where they are described by the Heisenberg $S$-matrix. It relates the two-body density matrix of both scattering particles before a collision $\hat\rho(1,2)$ with the one after a collision $\hat\rho(1,2)'$. Here $(1,2)$ label the quantum numbers of particles 1 and 2 in first quantization. We obtain
\begin{equation} 
\hat\rho(1,2)'=\hat{\mathcal{S}}\hat\rho(1,2)\hat{\mathcal{S}}^\dagger.
\end{equation}
In order to arrive at a single-particle description we trace out particle 2 later. We next assume that particles involved in a collision are uncorrelated, $\hat\rho(1,2)=\hat\rho(1)\otimes\hat\rho(2)$, both before and after the collision, an assumption justified for a system with a large number of particles. This assumption in fact corresponds to Boltzmann's original molecular chaos hypothesis (Stosszahlansatz). For the desired single-particle density matrices before and after a collision we obtain 
\begin{equation}
\hat\rho(1)=\frac12\mathrm{Tr}_2\lbrace(\openone-\mathcal P^\mathrm{ex})\hat\rho(1)\otimes\hat\rho(2)(\openone-\mathcal P^\mathrm{ex})\rbrace
\end{equation}
\begin{equation}
\hat\rho'(1)=\frac12\mathrm{Tr}_2\lbrace(\openone-\mathcal P^\mathrm{ex})\hat{\mathcal{S}}\hat\rho(1)\otimes\hat\rho(2)\hat{\mathcal{S}}^\dagger(\openone-\mathcal P^\mathrm{ex})\rbrace
\end{equation}
where we account for the indistinguishability of particles with the operator $\mathcal P^\mathrm{ex}$ exchanging the quantum numbers of particles 1 and 2. Due to fermionic statistics it comes with a minus sign.
This ansatz yields the following expression for the collision integral 
\begin{align}
\hat I_\mathrm{coll}\approx\frac{1}{\Delta t}\mathrm{Tr}_2\left\lbrace\frac{\openone-\mathcal P^\mathrm{ex}}{\sqrt2}\left[\hat{\mathcal{S}}\hat\rho(1)\otimes\hat\rho(2)\hat{\mathcal{S}}^\dagger\right.\right.\nonumber\\*
\left.\left.-\hat\rho(1)\otimes\hat\rho(2)\right]\frac{\openone-\mathcal P^\mathrm{ex}}{\sqrt2}\right\rbrace.
\label{eq:6}
\end{align}
The $S$-matrix is related to the $T$-matrix via $\hat{\mathcal{S}}=\openone-2\pi i \hat{\mathcal{T}}$ such that Eq.~(\ref{eq:6}) becomes
\begin{align}
\hat I_\mathrm{coll}\approx\frac{2\pi}{\Delta t}\mathrm{Tr}_2&\!\left\lbrace\frac{\openone\!-\!\mathcal P^\mathrm{ex}}{\sqrt2}\left[i \hat{\mathcal{T}}\hat\rho(1)\otimes\hat\rho(2)-i\hat\rho(1)\otimes\hat\rho(2)\hat{\mathcal{T}}^\dagger\right.\right.\nonumber\\*
&\left.\left.+2\pi\hat{\mathcal{T}}\hat\rho(1)\otimes\hat\rho(2)\hat{\mathcal{T}}^\dagger\right]\frac{\openone\!-\!\mathcal P^\mathrm{ex}}{\sqrt2}\right\rbrace.
\label{eq:9}
\end{align}
This expression then has to be evaluated in the phase-space representation. Before performing the trace operation, we compute the two-body Wigner transform of the expression in braces in Eq.~(\ref{eq:9}). This is a very lengthy exercise we demonstrate in detail in Appendix \ref{app:B}, as well as the subsequent trace, shown in Appendix \ref{app:C}. In the course of these calculations, we require the elements of the $S$-matrix. In the center-of-mass system they are given by
\begin{align}
&\langle 1:k,a;2:-k, b|\hat{\mathcal{T}}|1:k',c;2:-k',d\rangle=\nonumber\\*
&-2\pi i\delta\left(\epsilon_k-\epsilon_{k'}+Q_{abcd}\right)\nonumber\\*
&\times\langle1:a,2:b|\hat T(k,k')|1:c,2:d\rangle,
\label{eq:7}
\end{align}
where $k,k'$ denote the incoming and outgoing wave-vectors of the particles, and $m,n;m',n'$ incoming and outgoing spins respectively. The delta-function assures energy conservation, $\epsilon_k=\frac{\hbar^2 k^2}{2\mu}$ denotes kinetic energy with reduced mass $\mu=\frac{M}{2}$ and $Q_{abcd}\equiv Q(a^2+b^2-c^2-d^2)$ the shift in the quadratic Zeeman energy induced by a spin-changing collision. The on-shell $T$-matrix $\hat T(k,k')$ depends formally on the relative wave-vectors $k,k'$, but for our case of s-wave scattering, the dependence is only on the modulus. As they are related by energy conservation $|k'|=\sqrt{k^2+Q_{abcd}}$, effectively the dependence is only on $k$ or $k'$.
The QZE-shift vanishes for spin-conserving collisions, hence it is absent in the spin 1/2 case and has the effect that scattering processes with a large $Q_{abcd}$ are suppressed, because the $T$-matrix decays $\sim 1/|k'|$ for large $|k'|$. For ${}^{40}\mathrm K$, the splitting is given by the quadratic part of the Breit-Rabi-Formula \cite{BreitRabi}, $Q=-\frac{2\mu_\text{B}^2(g_J-g_I)^2B^2}{9^3a_\text{hfs}}$, with the Bohr magneton $\mu_\text{B}$, nuclear and electronic g-factors $g_I$, $g_J$ and hyperfine structure coefficient $a_\text{hfs}$ \cite{Arimondo1977}.

To account for the spin-dependent interactions we separate the $T$-matrix into channels of total spin $S$ and magnetization $M$ and obtain elements
\begin{align}
\langle1:a,2:b|&\hat T(k,k')|1:c,2:d\rangle\equiv T_{abcd}(k,k')\nonumber\\*
&=\sum_{SM}T_S(k,k')\left\langle ab|SM\right\rangle\left\langle SM|cd\right\rangle.
\end{align}
For a 1D system, the expression for a $T$-matrix in the channel with a coupling constant $g_S$ is
\begin{equation}
 T_S(k,k')=\frac{1}{2\pi}\frac{ik'\frac{2\hbar^2}{M}}{1-ik'\frac{2\hbar^2}{Mg_S} }.
\end{equation}
In each scattering channel we perform a low energy expansion in the coupling constant $g_S\rightarrow0$ up to second order, to maintain the unitarity of the $S$-matrix. Since for the 1D case, the expansion in powers of $g_S$ is accompanied by factors of $(k')^{-1}$, we artificially create a singularity in the imaginary part of the $T$-matrix in this procedure. We remedy this problem by choosing a cutoff $|k'|<\frac{Mg_S}{2\hbar^2}$, which is the distance between zero and the maximum of the imaginary part of the $T$-matrix (see Appendix \ref{app:D} for more details). This step is unnecessary in the 3D case discussed by Lhuillier and Lalo\"e \cite{Lhuillier1982I,Lhuillier1982II,Fuchs2003}. The result is then given by
\begin{align}
T_S(k,k')&\approx\frac{g_S}{2\pi}-\begin{cases} 0 & \mbox{if } |k'|\mbox{ $<\frac{Mg_S}{2\hbar^2}$} \\ \frac{iMg_S^2}{4\pi\hbar^2 k'}+\ldots & \mbox{if } |k'|\mbox{ $\geq \frac{Mg_S}{2\hbar^2}$ } \end{cases}
\end{align}
or respectively 
\begin{align}
\label{eq:23}
T_{abcd}(k,k')\approx\frac{U_{acbd}}{2\pi}-\begin{cases} 0 & \mbox{if } |k'|\mbox{ $<\frac{MU_{acbd}}{2\hbar^2}$} \\ \frac{i M\tilde U_{acbd}}{4\pi\hbar^2 k'}+\ldots & \mbox{if } |k'|\mbox{ $\geq\frac{MU_{acbd}}{2\hbar^2}$} \end{cases}.
\end{align}
The leading terms linear in $g_S$ correspond to forward scattering, the quadratic terms to backward (lateral in higher dimensions) scattering processes. We do not explicitly denote the spin-dependent cutoff in equations (\ref{eq:5}),(\ref{eq:26}) and (\ref{eq:27}), where it is used in the integrals over $q$.

The expansion of the $T$-matrix is performed in addition to a semiclassical gradient expansion of the Wigner function (see Appendix \ref{app:E} for details) to first order in the linear terms and to zero order in the quadratic expressions. During this procedure we encounter squares of delta functions whose interpretation is described in Appendix \ref{app:F}. Finally we obtain a collisional integral, consisting of three parts
\begin{equation}
I^\mathrm{coll}_{mn}(x,p)=I^\text{mf}_{mn}(x,p)+I^T_{mn}(x,p)+I^{T^2}_{mn}(x,p),
\label{eq:12}
\end{equation}
where $I^\text{mf}$ is linear in $a_S$ and contains the forward scattering part of collisions leading to phase-shifts,
\begin{align}
&I^\text{mf}_{mn}(x,p)\!=\!-\frac{i}{\hbar}\sum_l\!\!\left[V^\text{mf}_{nl}(x)W_{ml}(x,p)\!-\!W_{lm}(x,p)V^\text{mf}_{ln}(x)\right]\nonumber\\*
&\!+\!\frac12\sum_l\!\!\left\lbrace\partial_x V^\text{mf}_{nl}(x)\partial_pW_{ml}(x,p)\!+\!\partial_p W_{lm}(x,p)\partial_x V^\text{mf}_{ln}(x)\right\rbrace.
\end{align}
It coincides with the interaction term obtained from the treatment of the same problem on a simpler mean-field level \cite{Natu2009,Ebling2011}. Due to its effect as a non-linear modification of the trap and magnetic field, in Eq.~(\ref{eq:4}) we have separated this term from the collisional integral and added it to the free motion of the particles in the external fields.
The quadratic terms, which form Eq.~(\ref{eq:5}) contain backward scattering, including momentum exchange between particles. They appear as dissipation on the single-particle level and are given by
\begin{widetext}
\begin{align} 
\label{eq:26}
I_{mn}^{T}(x,p)&=-\frac{M}{\hbar^2}\sum_{abl}\left[\int_{q^2>\epsilon_1}\!\!\! dq\frac{\tilde U_{malb}}{\sqrt{q^2+\Delta_{mlab}}}W_{an}(x,p)W_{bl}(x,p-q)+\int_{q^2>\epsilon_2}\!\!\! dq\frac{\tilde U_{nalb}}{\sqrt{q^2+\Delta_{nlab}}}W_{ma}(x,p)W_{lb}(x,p-q)\right]\\*
\label{eq:27}
I_{mn}^{T^2}(x,p)&=\frac{M}{\hbar^2}\sum_{abcdl}\int_{q^2>\epsilon_3}\!\!\! dq\frac{U_{malb} U_{ncld}}{\sqrt{q^2+\Delta_{mnlabcd}}} W_{ac}(x,p-\tfrac12(q-\sqrt{q^2+\Delta_{mnlabcd}}))W_{bd}(x,p-\tfrac12(q+\sqrt{q^2+\Delta_{mnlabcd}})).
\end{align}
The integration domain cutoffs around $q=0$, coming from Eq.~(\ref{eq:23}) are given by $\epsilon_1=\frac{MU_{malb}}{\hbar}-\Delta_{mlab}$, $\epsilon_2=\frac{MU_{nalb}}{\hbar}-\Delta_{nlab}$ and $\epsilon_3=\frac{MU_{nalb}}{2\hbar}(U_{malb}+U_{ncld})-\Delta_{mnlabcd}$.
The corresponding result for a full 3D calculation (see \cite{Fuchs2003} for the case of spin 1/2) reads
\begin{align}
\label{eq:28}
&\frac{d}{dt}W_{mn}(\bm r,\bm p)+\left[\frac{\bm p}{M}\cdot\nabla_r-M\left(\omega_x^2x,\omega_y^2y,\omega_z^2z\right)\cdot\nabla_p+\frac{iQ}{\hbar}(n^2\!-m^2)\right]W_{mn}(\bm r,\bm p)\nonumber\\*
&+\frac{i}{\hbar}\sum_l\left[V^\text{mf}_{nl}(\bm r)W_{ml}(\bm r,\bm p)-W_{lm}(\bm r,\bm p)V^\text{mf}_{ln}(\bm r)\right]\nonumber\\*
 &-\frac12\sum_l\left\lbrace\nabla_r V^\text{mf}_{nl}(\bm r)\cdot\nabla_pW_{ml}(\bm r,\bm p)+\nabla_p W_{lm}(\bm r,\bm p)\cdot\nabla_r V^\text{mf}_{ln}(\bm r)\right\rbrace=I_{mn}^\text{coll}(\bm r,\bm p).
\end{align}
The mean-field potential is given by $V_{mn}^\text{mf}(\bm r)=2\int d\bm p\sum_{ab}U'_{mnab}W_{ab}(\bm r,\bm p)$
and the collision term reads 
\begin{align}
\label{eq:29}
I^\text{coll}_{mn}(\bm r,\bm p)&=-\frac{M}{4\pi\hbar^4}\int d\bm q\left\lbrace\sum_{abc}\left(\sqrt{q^2+\Delta_{mcab}}\tilde U'_{macb} W_{an}(\bm r,\bm p)W_{bc}(\bm r,\bm p-\bm q)\right.\right.\nonumber\\*
&\hspace{6cm}\left.\left.+\sqrt{q^2+\Delta_{abnc}}\tilde U'_{anbc}W_{ma}(\bm r,\bm p)W_{cb}(\bm r,\bm p-\bm q)\right)\right.\nonumber\\*
&-\left.\frac{1}{2\pi}\int d\Omega\sum_{abcdl}\sqrt{q^2+\Delta_{mnlabcd}}U'_{malb}U'_{ncld}W_{ac}(\bm r,\bm p-\tfrac12(\bm{q}-\bm p')W_{bd}(\bm r,\bm p-\tfrac12(\bm q+\bm p')\right\rbrace
\end{align}
where $\bm p'=\bm e_\Omega\sqrt{q^2+\Delta_{mnlabcd}}$ and $\bm e_\Omega$ denotes the unit vector corresponding to solid angle $d\Omega$.
\end{widetext}

A physical interpretation of this expression can be obtained by looking at the origin of the individual terms. The upper two lines of Eq.~(\ref{eq:29}) originate from the second order of the expansion of the $T$-matrix (\ref{eq:23}), describing the intensity shift in the forward scattered wave \cite{Lhuillier1982I,Fuchs2003}, while the first order only describes a phase shift and appears in the mean-field terms in (\ref{eq:28}). The bottom line of (\ref{eq:29}) contains all lateral scattering processes, hence the explicit angular dependence. In our formalism, the coupling constants $U$, $\tilde U$ include all particle indistinguishability and exchange contributions, which are discussed in greater detail in \cite{Lhuillier1982I}.

\section{Conclusion}
\label{sec5}
We have presented a novel approach to study relaxation dynamics in a closed quantum system, exploiting the unique properties of a large-spin Fermi sea. For this system, we have derived a multicomponent kinetic equation without phenomenological assumptions nor prior knowledge of the equilibrium state. As a key result, we find that this approach is well suited for the quantitative description of weakly interacting fermionic many-body systems with large spin. Both, the comparison of numerical simulations with full spatial resolution to a 1D experiment as well as the comparison of a simplified single-mode approximation to a 3D experiment yield a very good agreement without free parameters. We identify different collisional processes on different time scales and identify spin relaxation as the slowest dynamical process of the system. A variation of the density and the geometry of the system changes the respective spin relaxation rates by several orders of magnitude, ranging from a few milliseconds to several seconds. By tuning the magnetic field, we can precisely control the coupling strengths of individual collision channels, allowing to tune the character of a subsystem of two spin components within the large-spin Fermi sea continuously from an open to a closed system. The spin relaxation manifests itself in a perforation of the Fermi sea accompanied with a temperature increase. 

Our results broaden the understanding of many-body relaxation dynamics. In particular, the fermionic character of the system underlines its model character for various systems in nature. The possibility to monitor different spin components individually allows to employ the large-spin Fermi sea for novel studies of decoherence and relaxation processes in quantum many-body systems. Furthermore, spin relaxation dynamics might play an important role for proposed fermionic large-spin phenomena, e.g. quantum-chromodynamic-like color superfluidity or large-spin texture formation.

\begin{acknowledgments}
We acknowledge fruitful discussions with Jannes Heinze and Ludwig Mathey.
This work has been funded by Spanish Ministerio de Ciencia e Innovaci\'on (FIS 2008-00784, AAII-Hubbard, FPI-fellowship), ERC grants QUAGATUA, OSYRIS, EU IP SIQS and
Deutsche Forschungsgemeinschaft (DFG) grant FOR 801 and DFG excellence cluster The Hamburg Centre for Ultrafast Imaging - Structure, Dynamics, and Control of Matter on the Atomic Scale. 
\end{acknowledgments}

\appendix
\section{Experimental details}
\label{app:A}
\subsection{Preparation}
\label{app:A1}
We sympathetically cool spin-polarized \textsuperscript{40}K atoms in the state $\left|F=9/2,m=9/2\right\rangle$ down to a temperature of typically $0.1\,T_\text{F}$ in a magnetic trap, using bosonic \textsuperscript{87}Rb as a buffer gas. Subsequently we transfer the atoms into a crossed circular-elliptical optical dipole trap operated at a wavelength of $\lambda\,=\,812\,\text{nm}$. Using radio-frequency (rf) pulses and rf-sweeps, we create a spin mixture, which we evaporate to quantum degeneracy by lowering the power of the dipole trap exponentially in $2\,\text{s}$. This results in a sample with particle numbers of the order of $N\sim10^5$ at temperatures of $T=0.1\,-\,0.2\,T_\text{F}$. After the evaporation we compress the trap again to avoid particle loss during the experiments, realizing typical trapping frequencies of $\vec\omega=2\pi\times(33,33,137)\,\text{Hz}$. By varying the evaporation sequence and including additional waiting times, we can control the initial temperature and particle number independently in the same trap geometry. This allows us to modify the density while keeping $T/T_\text{F}$ approximately constant. Typically, a balanced mixture of atoms in spin states $m=\pm1/2$ is used as initial state throughout this paper. To study the spin-changing dynamics, we switch the magnetic field to low values. In FIG.~\ref{fig2}, beyond this, a coherent superposition of several spin states is prepared by applying subsequently a rf-pulse at low magnetic field corresponding to a spin rotation of $\theta=0.44$ (see \cite{Krauser2014} for more details). The 1D configuration used in FIG.~\ref{fig4} (a) is realized by adiabatically ramping up a 2D optical square lattice over $150\,\text{ms}$. The lattice is created by two orthogonal retro-reflected laser beams at wavelength $\lambda=1030\,\text{nm}$ with a $1/e^2$ radius of $200\,\mu\text{m}$ detuned with respect to each other by several tens of megahertz. The lattice depth is $25\,E_\text{recoil}$ with $E_\text{recoil}=\frac{\hbar^2k_\text{L}^2}{2M}$, where $k_\text{L}=\frac{2\pi}{\lambda}$. This creates an array of 1D tubes, where a single tube can be described as a harmonically trapped system with frequencies $\omega_x=2\pi\times84\,\text{Hz}$ and $\omega_{y,z}=2\pi\times 47\,\text{kHz}$. With a particle number of $N\approx100$ and $E_\text{F}=2\pi\hbar\times37\,\text{kHz}$, the radial trapping frequencies fulfill $\hbar\omega_{y,z}>E_\text{F}$ and at a temperature $k_\text{B}T=0.2\,E_\text{F}$, we can neglect a possible population of excited radial modes, hence we create a true 1D system. The extension of the radial ground state is around
1378 Bohr radii and thus one order of magnitude larger than the scattering lengths \cite{Krauser2012}. We thus neglect the possibility
of a confinement-induced resonance \cite{Olshanii1998}, and use the effective coupling constants $g_S^\text{1D}= 2\hbar\sqrt{\omega_y\omega_z}a_S$.

\subsection{Measurement}
\label{app:A2}
The relative populations of spin components are measured as follows: We release the atoms from the trap in an inhomogeneous magnetic field to separate the spin components during a time-of-flight expansion of typically $18.5\,\text{ms}$. We count the number of atoms in each spin component with resonant absorption imaging. For comparison, we measure the total number of particles as well as the temperature independently without the Stern-Gerlach field to avoid distortions of the particle cloud during the time-of-flight. The numbers given in this paper correspond to the initial temperature and particle number. In FIG.~\ref{fig7}(a), in order to extract the change in temperature over time, we determine the temperature only in one spin component, circumventing deviations associated with the imbalance of the spin mixture. For instance, to measure the temperature in $m=1/2$ we apply a sequence of linearly polarized microwave pulses with a duration of $50\,\mu s$ to transfer all significantly occupied spin components $m \neq 1/2$ into the $F=7/2$ hyperfine manifold of \textsuperscript{40}K. In the other hyperfine manifold the atoms are not resonant with the detection light and are thus obscured during the absorption imaging process.

\begin{widetext}
\section{Two-body Wigner transform}
\label{app:B}
Because the $T$-matrix depends only on the relative wave vectors we evaluate Eq.~(\ref{eq:9}) in the center-of-mass frame. We introduce the notation
\begin{align}
 R&=\frac12(x_1+x_2),\qquad  r=x_1-x_2,\nonumber\\*
   P&=p_1+p_2,\qquad  p=\frac12(p_1-p_2),
 \label{eq:a2}
\end{align} 
to denote center-of-mass and relative positions and momenta versus the coordinates of particles 1 and 2 denoted by subscript. We denote by $W^{(T,T^2)}$ the two-body Wigner transform of the part of Eq.~(\ref{eq:9}) linear in the $T$-matrix $2\pi i\frac{\openone-\mathcal P^\mathrm{ex}}{\sqrt2}\left[ \hat{\mathcal{T}}\hat\rho(1)\otimes\hat\rho(2)-\hat\rho(1)\otimes\hat\rho(2)\hat{\mathcal{T}}^\dagger\right]\frac{\openone-\mathcal P^\mathrm{ex}}{\sqrt2}$ and the quadratic part $4\pi^2\frac{\openone-\mathcal P^\mathrm{ex}}{\sqrt2}\left[\hat{\mathcal{T}}\hat\rho(1)\otimes\hat\rho(2)\hat{\mathcal{T}}^\dagger\right]\frac{\openone-\mathcal P^\mathrm{ex}}{\sqrt2}$ respectively. We obtain
\begin{equation}
W_{ijmn}^T(r,R,p,P)=\frac{-i}{2\pi\hbar^2}\int dK \int d\kappa e^{iKR} e^{i\kappa r}\langle K_{+},k_{+},i,m |\frac{\openone-\mathcal P^\mathrm{ex}}{\sqrt2}\hat{\mathcal{T}}\hat\rho(1)\otimes\hat\rho(2)\frac{\openone-\mathcal P^\mathrm{ex}}{\sqrt2}|K_{-},k_{-},j,n\rangle+h.c.
\end{equation}
\begin{equation}
W_{ijmn}^{T^2}(r,R,p,P)=\frac{1}{\hbar^2}\int dK \int d\kappa e^{iKR} e^{i\kappa r}\langle K_{+},k_{+},i,m|\frac{\openone-\mathcal P^\mathrm{ex}}{\sqrt2}\hat{\mathcal{T}}\hat\rho(1)\otimes\hat\rho(2)\hat{\mathcal{T}}^\dagger\frac{\openone-\mathcal P^\mathrm{ex}}{\sqrt2}|K_{-},k_{-},j,n\rangle
\end{equation}
where we introduced the wave-vectors $K_\pm=\frac{P}{\hbar}\pm\frac{K}2$ and $k_\pm=\frac{p}\hbar\pm\frac{\kappa}2$. We insert two complete bases $\int dK_1\int dk_1\sum_{ab}|K_1,k_1,a,b\rangle\langle K_1,k_1,a,b|$ and $\int dK_2\int dk_2\sum_{cd}|K_2,k_2,c,d\rangle\langle K_2,k_2,c,d|$ to the left and right of the tensor product of density matrices. The dependence of the $T$-matrix on the relative wave-vector only makes the integration over $K_{1,2}$ trivial. We substitute from (\ref{eq:7}) the expression
\begin{equation}
\langle K_1,k_1,a,b|\hat{\mathcal{T}}|K_2,k_2,c,d\rangle=\delta\left(\epsilon_{k_1}-\epsilon_{k_2}+Q_{abcd}\right)T_{abcd}(k_1,k_2)
\end{equation}
for the elements of the $T$-matrix into above expressions and obtain 
\begin{align}
W_{ijmn}^{T}&(r,R,p,P)=\frac{-i}{2\pi\hbar^2}\int dK\int d\kappa\int dk_1\int dk_2 e^{iKR} e^{i\kappa r}\sum_{abcd}\langle K_{+},k_{+},i,m|\frac{\openone-\mathcal P^\mathrm{ex}}{\sqrt2}\hat{\mathcal{T}}|K_{+},k_1,a,b\rangle\nonumber\\*
&\times\langle K_{+},k_1,a,b|\hat\rho(1)\otimes\hat\rho(2)|K_{-},k_2,c,d\rangle\langle K_{-},k_2,c,d|\frac{\openone-\mathcal P^\mathrm{ex}}{\sqrt2}|K_{-},k_{-},j,n\rangle+h.c.\nonumber\\*
&=\frac{-i}{4\pi\hbar^2}\int dK\int d\kappa\int dk_1\int dk_2e^{iKR} e^{i\kappa r}
\sum_{abcd}\delta(\epsilon_{k_{+}}-\epsilon_{k_1}+Q_{imab})\left(\delta(k_2-k_{-})\delta_{cj}\delta_{dn}-\delta(k_2+k_{-})\delta_{nc}\delta_{jd}\right)\nonumber\\*
&\times\left(T_{imab}(k_{+},k_1)-T_{miab}(-k_{+},k_1)\right)\langle K_{+},k_1,a,b|\hat\rho(1)\otimes\hat\rho(2)|K_{-},k_2,c,d\rangle+h.c.
\end{align}
for the linear term and
\begin{align}
W_{ijmn}^{T^2}&(r,R,p,P)=\frac{1}{2\hbar^2}\int dK\int d\kappa\int dk_1\int dk_2 e^{iKR} e^{i\kappa r}
\sum_{abcd}\delta(\epsilon_{k_{+}}-\epsilon_{k_1}+Q_{imab})\delta(\epsilon_{k_2}-\epsilon_{k_{-}}+Q_{cdjn})\nonumber\\*
&\times\left(T_{imab}(k_{+},k_1)-T_{miab}(-k_{+},k_1)\right)\left(T_{jncd}^{*}(k_{-},k_2)-T_{njcd}^{*}(-k_{-},k_2)\right)\langle K_{+},k_1,a,b|\hat\rho(1)\otimes\hat\rho(2)|K_{-},k_2,c,d\rangle
\end{align}
for the term quadratic in the $T$-matrix. The elements of the tensor product of density matrices are obtained from the Wigner functions by an inverse Wigner transform
\begin{align}
\langle K_{+},k_1,a,b|\hat\rho(1)\otimes\hat\rho(2)|K_{-},k_2,c,d\rangle=\hbar^2\int dR'\int dr' e^{-iKR'}e^{i(k_2-k_1)r'}\nonumber\\*
\times W_{ac}(R'+\tfrac{r'}{2},\tfrac{P+\hbar k_1+\hbar k_2}{2})W_{bd}(R'-\tfrac{r'}{2},\tfrac{P-\hbar k_1-k_2}{2}).
\end{align}
and we substitute this expression into the collision term. This produces a delta function $\int dK e^{iK(R-R')}=2\pi\delta(R-R')$ and after carrying out the integration over $K$ and $R'$ we obtain
\begin{align}
&W_{ijmn}^{T}(r,R,p,P)=\frac{-i}{2}\int d\kappa\int dk_1\int dk_2\int dr' e^{i\kappa r}e^{i(k_2-k_1)r'}\sum_{abcd}\left(\delta(k_2-k_{-})\delta_{cj}\delta_{dn}-\delta(k_2+k_{-})\delta_{nc}\delta_{jd}\right)\nonumber\\*
&\times\delta(\epsilon_{k_{+}}-\epsilon_{k_1}+Q_{imab})\left(T_{imab}(k_{+},k_1)-T_{miab}(-k_{+},k_1)\right)W_{ac}(R+\tfrac{r'}{2},\tfrac{P+\hbar k_1+\hbar k_2}{2})W_{bd}(R-\tfrac{r'}{2},\tfrac{P-\hbar k_1-\hbar k_2}{2})+h.c.
\end{align}
and
\begin{align}
&W_{ijmn}^{T^2}(r,R,p,P)=\pi\int d\kappa\int dk_1\int dk_2\int dr' e^{i\kappa r}e^{i(k_2-k_1)r'}\sum_{abcd}\delta(\epsilon_{k_{+}}-\epsilon_{k_1}+Q_{imab})\delta(\epsilon_{k_2}-\epsilon_{k_{-}}+Q_{cdjn})\nonumber\\*
&\times\left(T_{imab}(k_{+},k_1)-T_{miab}(-k_{+},k_1)\right)\left(T_{jncd}^{*}(k_{-},k_2)-T_{lncd}^{*}(-k_{-},k_2)\right)W_{ac}(R+\tfrac{r'}{2},\tfrac{P+\hbar k_1+\hbar k_2}{2})W_{bd}(R-\tfrac{r'}{2},\tfrac{P-\hbar k_1-\hbar k_2}{2})
\label{eq:a8}
\end{align}

\section{Trace over second particle}
\label{app:C}
In order to trace out particle 2 as described in (\ref{eq:9}) we return from the center-of-mass frame to he lab frame by substituting equations (\ref{eq:a2}) back into (\ref{eq:a8}). The trace over particle 2 means performing the operation $ I_{ij}^{(T,T^2)}(x_1,p_1)=\frac{1}{\Delta t}\int dx_2\int dp_2\sum_{mn}\delta_{mn} W_{ijmn}^{(T,T^2)}(r,R,p,P)$
on each term. Introducing the notations $q=2\hbar k$, $p_1'=p_1-\tfrac{q-\hbar(k_1+k_1)}2$ and $p_2'=p_1-\tfrac{q+\hbar(k_1+k_2)}2$ we arrive at the following expressions for the collision term:
\begin{align}
&I_{ij}^{T}(x_1,p_1)=\frac{-i}{2\Delta t}\int d\kappa\int dk_1\int dk_2\int dr'\int dr\int dq e^{i\kappa r}e^{i(k_2-k_1)r'}\sum_{abcdl}\delta(\epsilon_{k_{+}}-\epsilon_{k_1}+Q_{ilab})\nonumber\\*
&\left(\delta(k_2-k_{-})\delta_{cj}\delta_{dl}-\delta(k_2+k_{-})\delta_{lc}\delta_{jd}\right) \left(T_{ilab}(k_{+},k_1)-T_{liab}(-k_{+},k_1)\right)W_{ac}(x_1-\tfrac{r-r'}2,p_1')W_{bd}(x_1-\tfrac{r+r'}{2},p_2')+h.c.
\end{align}
and
\begin{align}
&I_{ij}^{T^2}(x_1,p_1)=\frac{\pi}{\Delta t}\int d\kappa\int dk_1\int dk_2\int dr'\int dr\int dq e^{i\kappa r}e^{i(k_2-k_1)r'}\sum_{abcdl}\delta(\epsilon_{k_{+}}-\epsilon_{k_1}+Q_{ilab})\delta(\epsilon_{k_2}-\epsilon_{k_{-}}+Q_{cdjl})\nonumber\\*
&\times\left(T_{ilab}(k_{+},k_1)-T_{liab}(-k_{+},k_1)\right)\left(T_{jlcd}^{*}(k_{-},k_2)-T_{ljcd}^{*}(-k_{-},k_2)\right)W_{ac}(x_1-\tfrac{r-r'}2,p_1')W_{bd}(x_1-\tfrac{r+r'}{2},p_2').
\label{eq:a9}
\end{align}
\end{widetext}

\section{One-dimensional large-spin T-matrix}
\label{app:D}
In the one-dimensional two-body scattering problem in the center-of-mass frame with Hamiltonian $H=\frac{-\hbar^2}{2\mu}\frac{\text d^2}{\text dx^2}+g_S\delta(x)$ the wave function is $\psi(x)=e^{ikx}+f_{k'}e^{ik'|x|}$, from which follows $f_{k'=}\frac{-1}{1-i\hbar^2 k'/\mu g_S}$ for the scattering amplitude. The scattered wave function $\psi_\text{sc}(k')=f_{k'}e^{i{k'}|x|}$ and $T$-matrix are related through the Green's function $\psi_\text{sc}(k')=G(k,k')T(k',k)$, which in 1D is given by
\begin{equation}
G(x)=\frac{2\mu}{\hbar^2}\int dk'\frac{e^{ik'x}}{k^2+k'^2+i0^+}=2\pi\frac{i\mu}{\hbar^2 k}e^{ik|x|}
\end{equation}
such that
\begin{equation}
 T_S(k,k')=\frac{1}{2\pi}\frac{ik'\frac{2\hbar^2}{M}}{1-ik'\frac{2\hbar^2}{Mg_S} }.
 \label{eq:b4}
\end{equation}
In the presence of a quadratic Zeeman shift $Q$ there is a difference in modulus of incoming and outgoing wave-vectors $|k'|=\sqrt{k^2+Q}$. Here and throughout this paper, if the argument of the square root becomes negative for a negative $Q$ the $T$-matrix vanishes and with it the entire collision term.

\begin{figure}[t]
\centering
\includegraphics[width=8.6cm]{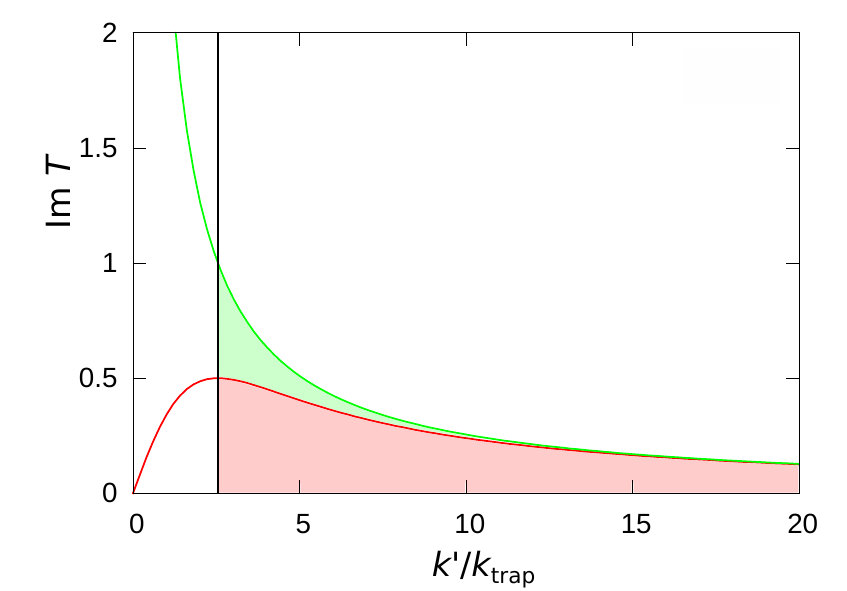}
\caption{Comparison of the imaginary part of the $T$-matrix (\ref{eq:b5}) (red) with the expansion (\ref{eq:b6}) (green) for a small coupling constant $g_S$. To avoid the singularity at $k'=0$ we choose $T_S=0$ inside the region $|k'|\leq \frac{Mg_S}{2\hbar^2}$ indicated by the black line. The wave vector is scaled in terms of the trapping frequency: $k_\text{trap}=\sqrt{M\omega/\hbar}$.}
\label{fig8}
\end{figure}

A problem absent in the 3D case is encountered during the low-energy expansion of (\ref{eq:b4}). The imaginary part of the $T$-matrix is given by
\begin{equation}
\text{Im}T_S(k,k')=\frac{1}{2\pi}\frac{k'\frac{2\hbar^2}{M}}{1+\frac{k'^24\hbar^4}{M^2 g_S^2}}
\label{eq:b5}
\end{equation}
and an expansion in powers of $g_S$ produces a singularity for $k'=0$, since
\begin{equation}
\text{Im}T_S(k,k')=\frac{1}{2\pi}\frac{g_S^2M}{2\hbar^2k'}+\ldots
\label{eq:b6}
\end{equation}
This singularity is artificial and we use a cutoff to circumvent it. We choose the cutoff to be the maximum of $\text{Im}T$ at $k'=\frac{Mg_S}{2\hbar^2}$, as depicted in FIG.~\ref{fig8}. So we use the expression 
\begin{align}
T_S(k,k')&\approx\frac{g_S}{2\pi}-\begin{cases} 0 & \mbox{if } |k'|\mbox{ $<\frac{Mg_S}{2\hbar^2}$} \\ \frac{iMg_S^2}{4\pi\hbar^2 k'}+\ldots & \mbox{if } |k'|\mbox{ $\geq \frac{Mg_S}{2\hbar^2}$ } \end{cases}.
\label{eq:b7}
\end{align}
to expand the $T$-matrix.

\section{Semiclassical gradient expansion}
\label{app:E}
In order to further simplify the expressions (\ref{eq:a9}) we assume the Wigner function to vary only slowly in space compared to single-particle wave-functions. This assumption means that local contributions to the collision term dominate and we perform a Taylor expansion for the spatial coordinate
\begin{equation}
W_{ij}(x_1-\tfrac{r\pm r'}2,p)= W_{ij}(x_1,p)-\tfrac{r\pm r'}{2}\partial_{x_1}W_{ij}(x_1,p)+\ldots
\end{equation}
therefore the expansion of the product of Wigner functions in (\ref{eq:a9}) reads
\begin{align}
&W_{ac}(x_1-\tfrac{r-r'}2,p_1')W_{bd}(x_1-\tfrac{r+r'}{2},p_2')=\nonumber\\*
&W_{ac}(x_1,p_1')W_{bd}(x_1,p_2')-\tfrac{r}{2}\partial_{x_1}\left(W_{ac}(x_1,p_1')W_{bd}(x_1,p_2')\right)\nonumber\\*
&+\tfrac{r'}{2}W_{bd}(x_1,p_2')\partial_{x_1}W_{ac}(x_1,p_1')\nonumber\\*
&-\tfrac{r'}{2}W_{ac}(x_1,p_1')\partial_{x_1}W_{bd}(x_1,p_2')+\ldots
\label{eq:gradient_expansion}
\end{align}
Together with the expansion of the $T$-matrix above we must be careful to expand the collision term in two small parameters in a meaningful way. One small parameter is the coupling constant proportional to the s-wave scattering length. The other one is related to the gradient expansion. Its magnitude is determined by the Fermi or thermal wavelength compared to the variation of the Wigner function determined by the system size. To maintain the unitarity of the $S$-matrix, we expand the $T$-matrix to second order. This means we will obtain terms linear in $a_S$ from $I^{T}_{ij}(x,p)$ and quadratic terms from $I^{T}_{ij}(x,p)$ and $I^{T^2}_{ij}(x,p)$. We expand the terms linear in $a_S$ up to first order in gradients and the terms quadratic in $a_S$ to zero order, keeping only the local term. This amount to a semi-classical approximation of the theory. In this case we substitute $W_{ac}(x_1-\tfrac{r-r'}2,p_1')W_{bd}(x_1-\tfrac{r+r'}{2},p_2')\approx W_{ac}(x_1,p_1')W_{bd}(x_1,p_2')$ into (\ref{eq:a9}), which means that further delta functions $\int dr'e^{i(k_2-k_1)r'}=2\pi\delta(k_2-k_1)$, $\int dr e^{i\kappa r}=2\pi\delta(\kappa) $ appear. We introduce renamed variables $k_\pm\rightarrow k$, $k_1\
\rightarrow k'$, $x_1,p_1\rightarrow x,p$ and $p_{\pm}\equiv p-\
\hbar(k\pm k')$ the local parts of the collision integral become
\begin{align}
&I_{ij}^{T}(x,p)=\frac{-i2\pi}{\Delta t}\int dq\int dk'\sum_{abcdl}\delta(\epsilon_k-\epsilon_{k'}+Q_{ilab})\nonumber\\*
&\times\left(\delta(k-k')\delta_{jc}\delta_{ld}-\delta(k+k')\delta_{lc}\delta_{jd}\right)\left(U_{ialb}-\frac{iM\tilde U_{ialb}}{\hbar^2 k'}\right)\nonumber\\*
&\times W_{ac}(x,p_{-})W_{bd}(x,p_{+})+h.c.
\end{align}
and
\begin{align}
&I_{ij}^{T^2}(x,p)=\frac{2\pi}{\Delta t}\int dq\int dk'\sum_{abcdl}\delta(\epsilon_k-\epsilon_{k'}+Q_{ilab})\nonumber\\*
&\times\delta(\epsilon_{k'}-\epsilon_{k}+Q_{cdjl}) U_{ialb} U_{jcld} W_{ac}(x,p_{-})W_{bd}(x,p_{+}).
\label{eq:collision2}
\end{align}

\section{Squares and products of delta functions}
\label{app:F}
In scattering theory, the square of a delta function of energy appears frequently, when terms quadratic in the $T$-matrix are involved. A well-known interpretation of this is $\left[\delta(E)\right]^2\approx\frac{\Delta t}{2\pi\hbar}\delta(E)$, where $\Delta t$ denotes the elapsed time interval, which is quasi-infinite when compared to the duration of a single scattering event but nevertheless short compared to other relevant dynamics, like relaxation or the trapping period. This approximation is obtained by using the Fourier representation of the delta function
\begin{equation}
\delta(E)=\frac{1}{2\pi\hbar}\int dt e^{\frac i\hbar Et}
\end{equation}
such that
\begin{align}
\left[\delta(E)\right]^2&=\delta(E)\frac{1}{2\pi\hbar}\int dt e^{\frac i\hbar Et}=\delta(E)\frac{1}{2\pi\hbar}\int dt\nonumber\\*
&\approx\delta(E)\frac{1}{2\pi\hbar}\int_{\Delta t} dt=\frac{\Delta t}{2\pi\hbar}\delta(E)
\end{align}
This can also be applied to products of the form $\delta(\epsilon_k-\epsilon_{k'})\delta(k-k')$ since
\begin{align}
\delta &\left(\frac{\hbar^2 k^2}{2\mu}-\frac{\hbar^2 k'^2}{2\mu}\right)\delta(k-k')=\nonumber\\
&=\frac{\mu}{\hbar^2|k'|}\delta(|k|-|k'|)\delta(|k|-|k'|)\delta_{\text{sgn}(k),\text{sgn}(k')}\nonumber\\
&=\frac{\hbar^2|k'|}{\mu}\delta_{\text{sgn}(k),\text{sgn}(k')}\left[\delta\left(\frac{\hbar^2 k^2}{2\mu}-\frac{\hbar^2 k'^2}{2\mu}\right)\right]^2\nonumber\\
&\approx\frac{\hbar|k'|\Delta t}{2\pi\mu}\delta_{\text{sgn}(k),\text{sgn}(k')}\delta\left(\frac{\hbar^2 k^2}{2\mu}-\frac{\hbar^2 k'^2}{2\mu}\right)\nonumber\\
&=\frac{\Delta t}{2\pi\hbar}\delta(k-k')
\label{eq:delta1}
\end{align}
We modify this approximation to take into account the shift $Q$ in the quadratic Zeeman energy after a spin-changing collision. In our calculations two situations appear. In the first, coming from (\ref{eq:delta1}), there is only one shift and we must be careful that only the delta-function with the shift comes from a $T$-matrix where we can approximate the integration area with the interval $\Delta t$: 
\begin{align}
&\delta(\epsilon_k-\epsilon_{k'})\delta(\epsilon_k-\epsilon_{k'}+Q)\nonumber\\
&\approx\delta(\epsilon_k-\epsilon_{k'})\frac{1}{2\pi\hbar}\int_{\Delta t} dt e^{\frac i\hbar(\epsilon_k-\epsilon_{k'}+Q)t}\nonumber\\
&=\delta(\epsilon_k-\epsilon_{k'})\frac{1}{2\pi\hbar}\int_{\Delta t} dt e^{\frac{i}{\hbar}Qt/}\nonumber\\
&=\frac{\Delta t}{2\pi\hbar}\delta(\epsilon_k-\epsilon_{k'})\text{sinc}\left(\frac{Q\Delta t}{2\hbar}\right).
\end{align}
In the second case, both delta-functions originate from the energy conservation of the $T$-matrix
\begin{align}
&\delta(\epsilon_k-\epsilon_{k'}+Q_1)\delta(\epsilon_k-\epsilon_{k'}+Q_2)\nonumber\\
&=\frac{1}{(2\pi\hbar)^2}\int_{\Delta t}\! dt\int_{\Delta t}\! dt' e^{\frac{i}{\hbar}(\epsilon_k-\epsilon_{k'}+Q_1)t}e^{\frac{i}{\hbar}(\epsilon_k-\epsilon_{k'}+Q_2)t'}\nonumber\\
&=\frac{1}{(2\pi\hbar)^2}\int_{\Delta t}\! dt\int_{\Delta t}\! dt'e^{\frac i\hbar((\epsilon_k-\epsilon_{k'})(t+t')+Q_1 t+Q_2 t')}\nonumber\\
&=\frac{2}{(2\pi\hbar)^2}\int_{\Delta t}\! du\int_{\Delta t}\! du'e^{\frac i\hbar(\epsilon_k-\epsilon_{k'})u}e^{\frac{iQ_1}{2\hbar} (u-u')}e^{\frac{iQ_2}{2\hbar}(u+u')}\nonumber\\
&=\frac{2}{(2\pi\hbar)^2}\int_{\Delta t}\! du\int_{\Delta t}\! du'e^{\frac i\hbar(\epsilon_k-\epsilon_{k'}+\frac{Q_1+Q_2}{2})u}e^{\frac{i}{2\hbar}(Q_2-Q_1)u'}\nonumber\\
&\approx\frac{\Delta t}{2\pi\hbar}\delta\left(\epsilon_k-\epsilon_{k'}+\tfrac12 (Q_1+Q_2)\right)\text{sinc}\left(\frac{Q_2-Q_1}{2\hbar}\Delta t\right).
\end{align}
The time interval $\Delta t$ that appears in front cancels with the one introduced at the beginning (\ref{eq:9}) and for the sinc-function we assume it to be small such that $\text{sinc}\rightarrow 1$.

\begin{widetext}
\section{Single-mode approximation with QZE and in 1D}
\label{app:G}
Taking the quadratic Zeeman effect into account, the expressions for the single-mode approximation (\ref{eq:sma}) become slightly more complicated. The equation of motion is now given by
\begin{equation}
\label{eq:g1}
 \frac{d}{dt}M_{mn}=-\frac{M}{4\pi\hbar^4}\left\lbrace\sum_{abl}\left(\lambda'^{(1)}_{mlab} \tilde U'_{malb}M_{an}M_{bl}+\lambda'^{(1)}_{nlab}\tilde U'_{nalb}M_{ma}M_{lb}\right)-\sum_{abcdl}\lambda'^{(2)}_{mnlabcd}U'_{malb} U'_{ncld} M_{ac}M_{bd}\right\rbrace,
\end{equation}
where the two now separate integrals $\lambda^{(1,2)}$ are spin-dependent and given by
\begin{equation}
 \lambda'^{(1)}_{abcd}=\frac1N\int d\bm r\int d\bm p\int d\bm q \sqrt{\bm q^2+\Delta_{abcd}}f_0(\bm r,\bm p)f_0(\bm r,\bm p-\bm q),
\end{equation}
 and
 \begin{align}
  \lambda'^{(2)}_{mnlabcd}=\frac1N\int d\bm r\int d\bm p\int d\bm q \sqrt{\bm q^2+\Delta_{mnlabcd}}f_0(\bm r,\bm p)f_0(\bm r,\bm p-\bm q),
 \end{align}
 respectively, where the infrared cutoff described in Appendix~\ref{app:D} must be employed.

 The single-mode approximation can also be applied to the 1D system, from the Boltzmann equation (\ref{eq:4}). Equation (\ref{eq:g1}) changes to
 \begin{equation}
  \frac{d}{dt}M_{mn}=-\frac{M}{\hbar^2}\left\lbrace\sum_{abl}\left(\lambda^{(1)}_{mlab} \tilde U_{malb}M_{an}M_{bl}+\lambda^{(1)}_{nlab}\tilde U_{nalb}M_{ma}M_{lb}\right)-\sum_{abcdl}\lambda^{(2)}_{mnlabcd}U_{malb} U_{ncld} M_{ac}M_{bd}\right\rbrace,
 \end{equation}
 and the other expressions become
 \begin{align}
  \lambda^{(1)}_{abcd}=\frac1N\int dx\int dp\int_{q^2>\epsilon_{1,2}} dq \frac{f_0(r,p)f_0(r,p-q)}{\sqrt{q^2+\Delta_{abcd}}},
 \end{align}
 and
 \begin{align}
  \lambda^{(2)}_{mnlabcd}=\frac1N\int dx\int dp\int_{q^2>\epsilon_3} dq \frac{f_0(r,p)f_0(r,p-q)}{\sqrt{q^2+\Delta_{mnlabcd}}},
 \end{align}
 respectively, with the equilibrium distribution
 \begin{equation}
  \label{eq:g7}
  f_0(x,p)=\left\lbrace\exp\left[\frac{1}{k_BT}\left(\frac{p^2}{2M}+\frac12 M\omega^2 x^2-\mu\right)\right]+1\right\rbrace^{-1}.
 \end{equation}
 \end{widetext}
  A comparison of 1D single-mode results with the full 1D Boltzmann equation shows good agreement for pure spin relaxation as shown in FIG.~\ref{fig4}.
  
  A further inclusion of the coherent oscillations described by the commutator in Eq.~(\ref{eq:4}) into the single-mode equation shows that while the oscillations themselves are reproduced with high accuracy \cite{Krauser2014}, the damping of coherent oscillations such as in FIG.~\ref{fig3} is not captured well. We attribute this to the fact that damping is driven by the much stronger spin-conserving collisions and affects more strongly the individual phase-space distributions of the spin states, which are taken to be constant in time in the single-mode approximation. Thus we consider it necessary to use the full 1D Boltzmann in these cases.

\begin{figure}[t]
\centering
\includegraphics[width=8.6cm]{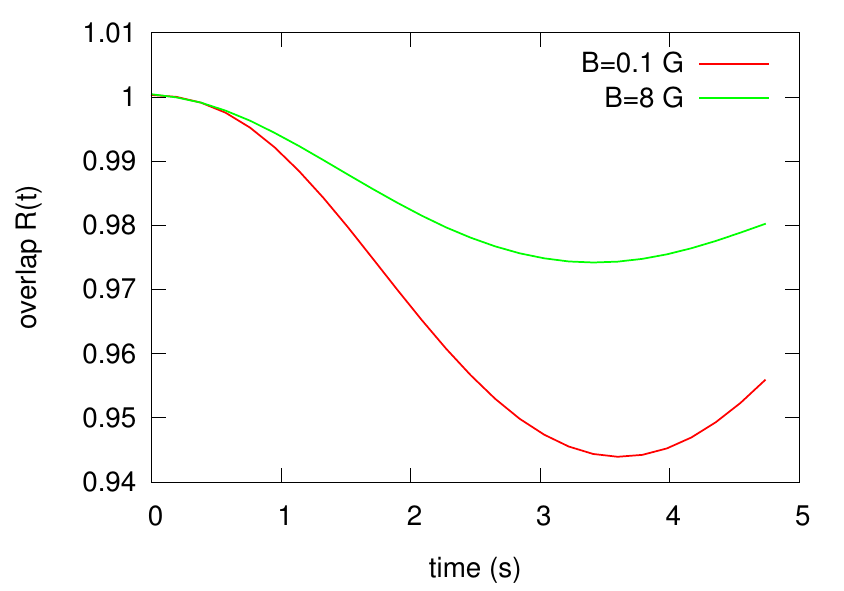}
\caption{Overlap between a non-interacting equilibrium distribution (\ref{eq:g7}) and the Wigner function during the simulations performed to obtain the temperatures in FIG.~\ref{fig7}(b).}
\label{fig9}
\end{figure}

\section{Concept of temperature in FIG. 7}
\label{app:H}
Under the assumptions stated at the end of section \ref{sec3}, we extract a temperature from our 1D numerical simulations as follows. At each time, we have the full Wigner function available, and can compare the Wigner-function of e.g. the $m=1/2$ component $W_{\frac 12\frac12}(x,p,t)\equiv W(x,p,t)$ to a non-interacting equilibrium distribution $f_0(x,p)$ (\ref{eq:g7}). This distribution is determined by particle number 
\begin{equation}
N=\int dx \int dp f_0(x,p),
\end{equation}
and trap energy
\begin{equation}
E=\int dx\int dp\left( \frac{p^2}{2M}+\frac12M\omega^2x^2\right) f_0(x,p),
\end{equation}
but also equivalently by temperature and chemical potential. Hence, we calculate at each time the particle number and trap energy of $W(x,p,t)$ and generate a Fermi distribution $f_0(x,p,t)=f_0(N(t),E(t))=f_0(\mu(t),T(t))$ with the same values for $N$ and $E$. The temperature of this distribution is plotted in FIG.~\ref{fig7}(b) as an estimate for the temperature of $W$. The overlap between this equilibrium distribution and the Wigner function, 
\begin{equation}
 R(t)=\frac{\int dx\int dp f_0(x,p,t)W_{\frac 12\frac12}(x,p,t)}{\int dx\int dp W_{\frac 12\frac12}(x,p,t)W_{\frac 12\frac12}(x,p,t)},
 \end{equation}
 is plotted in FIG.~\ref{fig9} and for the times we consider maintains sufficiently large values.


\begin{thebibliography}{50}
\bibitem{Srednicki1994}
M.~Srednicki,
\textit{Chaos and quantum thermalization},
Phys. Rev. E \textbf{50}, 888 (1994)

\bibitem{Rigol2008}
M.~Rigol, V.~Dunjko, and M.~Olshanii,
\textit{Thermalization and its mechanism for generic isolated quantum systems},
Nature (London) \textbf{452}, 854-858 (2008)

\bibitem{Dziarmaga2010}
J.~Dziarmaga,
\textit{Dynamics of a Quantum Phase Transition and Relaxation to a Steady State},
Adv. Phys. \textbf{59}, 6, 1063-1189 (2010)

\bibitem{Polkovnikov2011}
A.~Polkovnikov, K.~Sengupta, A.~Silva, and M.~Vengalattore,
\textit{Nonequilibrium dynamics of closed interacting quantum systems},
Rev. Mod. Phys. \textbf{83}, 863-883 (2011)

\bibitem{Gring2012}
M.~Gring, M.~Kuhnert, T.~Langen, T.~Kitagawa, B.~Rauer, M.~Schreitl, I.~Mazets, D.~Adu~Smith, E.~Demler, and J.~Schmiedmayer,
\textit{Relaxation and Prethermalization in an Isolated Quantum System},
Science \textbf{337}, 1318-1322 (2012)

\bibitem{Langen2013}
T.~Langen, R.~Geiger, M.~Kuhnert, B.~Rauer, and J.~Schmiedmayer,
\textit{Local emergence of thermal correlations in an isolated quantum many-body system},
Nat. Phys. \textbf{9}, 640-643 (2013)

\bibitem{Berges2004}
J.~Berges, Sz.~Bors\'anyi, and C.~Wetterich, 
\textit{Prethermalization},
Phys. Rev. Lett. \textbf{93}, 142002 (2004)

\bibitem{Kinoshita2006}
T.~Kinoshita, T.~Wenger, D.~S.~Weiss,
\textit{A quantum Newton's cradle},
Nature (London) \textbf{440}, 900 (2006)

\bibitem{Hofferberth2007}
S.~Hofferberth, I.~Lesanovsky, B.~Fischer, T.~Schumm, and J.~Schmiedmayer,
\textit{Non-equilibrium coherence dynamics in one-dimensional Bose gases},
Nature (London) \textbf{449}, 324-327 (2007)

\bibitem{Cheneau2012}
 M.~Cheneau, P.~Barmettler, D.~Poletti, M.~Endres, P.~Schau\ss, T.~Fukuhara, C.~Gross, I.~Bloch, C.~Kollath, and S.~Kuhr,
\textit{Light-cone-like spreading of correlations in a quantum many-body system},
Nature (London) \textbf{481}, 484-487 (2012)   

\bibitem{Trotzky2012}
S.~Trotzky, Y.-A.~Chen, A.~Flesch, I.~P.~McCulloch, U.~Schollw\"ock, J.~Eisert, and I. Bloch,
\textit{Probing the relaxation towards equilibrium in an isolated strongly correlated one-dimensional Bose gas},
Nat. Phys. \textbf{8}, 325-330 (2012)

\bibitem{Lux2013}
J.~Lux, J.~M\"uller, A.~Mitra, and A.~Rosch,
\textit{Hydrodynamic long-time tails after a quantum quench},
ArXiv 1311.7644 (2013) 

\bibitem{Erhard2004}
M.~Erhard, H.~Schmaljohann, J.~Kronj\"ager, K.~Bongs, and K.~Sengstock, 
\textit{Bose-Einstein condensation at constant temperature}, 
Phys. Rev. A \textbf{70}, 031602 (2004)

\bibitem{Sadler2006}
L.~E.~Sadler, J.~M.~Higbie, S.~R.~Leslie, M.~Vengalattore, and D.~M.~Stamper-Kurn,
\textit{Spontaneous symmetry breaking in a quenched ferromagnetic spinor Bose-Einstein condensate},
Nature (London) \textbf{443}, 312-315 (2006)

\bibitem{Griesmaier2005}
A.~Griesmaier, J.~Werner, S.~Hensler, J.~Stuhler, and T.~Pfau,
\textit{Bose-Einstein Condensation of Chromium},
Phys. Rev. Lett. \textbf{94}, 160401 (2005)

\bibitem{Schmaljohann2004}
H.~Schmaljohann, M.~Erhard, J.~Kronj\"ager, M.~Kottge, S.~van Staa, L.~Cacciapuoti, J.~J.~Arlt, K.~Bongs, and K.~Sengstock,
\textit{Dynamics of $F=2$ Spinor Bose-Einstein Condensates},
Phys. Rev. Lett. \textbf{92}, 040402 (2004)

\bibitem{Klempt2009}
C.~Klempt, O.~Topic, G.~Gebreyesus, M.~Scherer, T.~Henninger, P.~Hyllus, W.~Ertmer, L.~Santos, and J.~J.~Arlt, 
\textit{Multiresonant spinor dynamics in a Bose-Einstein condensate},
Phys. Rev. Lett. \textbf{103}, 195302 (2009)

\bibitem{Kronjaeger2010} 
J.~Kronj\"ager, C.~Becker, P.~Soltan-Panahi, K.~Bongs, and K.~Sengstock, 
\textit{Spontaneous Pattern Formation in an Antiferromagnetic Quantum Gas},
Phys. Rev. Lett. \textbf{105}, 090402 (2010)

\bibitem{MurPetit2006}
J.~Mur-Petit, M.~Guilleumas, A.~Polls, A.~Sanpera, M.~Lewenstein, K.~Bongs, and K.~Sengstock,
\textit{Dynamics of $F=1$ \textsuperscript{87}Rb condensates at finite temperatures},
Phys. Rev. A \textbf{73}, 013629 (2006)

\bibitem{Santos2006}
L.~Santos, and T.~Pfau,
\textit{Spin-3 Chromium Bose-Einstein Condensates},
Phys. Rev. Lett. \textbf{96}, 190404 (2006)

\bibitem{StamperKurn2012}
D.~M.~Stamper-Kurn, and M.~Ueda,
\textit{Spinor Bose gases: Explorations of symmetries, magnetism and quantum dynamics},
Rev. Mod. Phys. \textbf{85}, 1191-1244 (2013) 

\bibitem{Pechkis2013}
H.~K.~Pechkis, J.~P.~Wrubel, A.~Schwettmann, P.~F.~Griffin, R.~Barnett, E.~Tiesinga, and P.~D.~Lett,
\textit{Spinor dynamics in an antiferromagnetic spin-1 thermal Bose gas},
Phys. Rev. Lett. \textbf{111}, 025301 (2013)

\bibitem{Regal2004}
C.~A.~Regal, M.~Greiner, and D.~S.~Jin,
\textit{Observation of Resonance Condensation of Fermionic Atom Pairs},
Rev. Lett. \textbf{92}, 040403 (2004)

\bibitem{Zwierlein2005}
M.~W.~Zwierlein, J.~R.~Abo-Shaeer, A.~Schirotzek, C.~H.~Schunck, and W.~Ketterle,
\textit{Vortices and superfluidity in a strongly interacting Fermi gas},
Nature (London) \textbf{435}, 1047 (2005)

\bibitem{Sommer2011}
A.~Sommer, M.~Ku, G.~Roati, and M.~W.~Zwierlein,
\textit{Universal spin transport in a strongly interacting Fermi gas},
Nature (London) \textbf{472}, 201 (2011)

\bibitem{Koschorrek2013}
M. Koschorreck, D. Pertot, E. Vogt, and M. K\"ohl,
\textit{Universal spin dynamics in two-dimensional Fermi gases},
Nat. Phys. \textbf{9}, 405 (2013)

\bibitem{Du2008}
X.~Du, L.~Luo, B.~Clancy, and J.~E.~Thomas,
\textit{Observation of Anomalous Spin Segregation in a Trapped Fermi Gas},
Phys. Rev. Lett. \textbf{101}, 150401 (2008)

\bibitem{Natu2009}
S.~S.~Natu, and E.~J.~Mueller,
\textit{Anomalous spin segregation in a weakly interacting two-component Fermi gas},
Phys. Rev. A \textbf{79}, 051601(R) (2009)

\bibitem{Jo2009}
G.~B.~Jo, Y.~R.~Lee, J.~H.~Choil, C.~A.~Christensen, T.~H.~Kim, J.~H.~Thywissen, D.~E.~Pritchard, and W.~Ketterle, 
\textit{Itinerant Ferromagnetism in a Fermi Gas of Ultracold Atoms},
Science \textbf{325}, 1521-1524 (2009)

\bibitem{Conduit2011}
G.~J.~Conduit, and E.~Altman, 
\textit{Effect of three-body loss on itinerant ferromagnetism in an atomic Fermi gas},
Phys. Rev. A \textbf{83}, 043618 (2011)

\bibitem{Zhang2011}
S.~Zhang, and T.-L.~Ho,
\textit{ Atom loss maximum in ultra-cold Fermi gases},
New J. Phys. \textbf{13}, 055003 (2011)

\bibitem{Pekker2011}
D.~Pekker, M.~Babadi, R.~Sensarma, N.~Zinner, L.~Pollet, M.~W.~Zwierlein, and E.~Demler, 
\textit{Competition between pairing and ferromagnetic instabilities in ultracold Fermi gases near Feshbach resonances},
Phys. Rev. Lett. \textbf{106}, 050402 (2011)

\bibitem{Krauser2012}
J.~S.~Krauser, J.~Heinze, N.~Fl\"aschner, S.~G\"otze, O.~J\"urgensen, D.-S.~L\"uhmann, C.~Becker, and K.~Sengstock,
\textit{Coherent multi-flavour spin dynamics in a fermionic quantum gas},
Nat. Phys. \textbf{8}, 813 (2012)

\bibitem{Dong2013}
Y.~Dong, and H.~Pu,
\textit{Spin mixing in spinor Fermi gases},
Phys. Rev. A \textbf{87}, 043610 (2013)

\bibitem{Heinze2013}
J.~Heinze, J.~S.~Krauser, N.~Fl\"aschner, K.~Sengstock, C.~Becker, U.~Ebling, A.~Eckardt, and M.~Lewenstein,
\textit{Engineering spin-waves in a high-spin ultracold Fermi gas},
Phys. Rev. Lett. \textbf{110}, 250402 (2013)

\bibitem{Krauser2014}
J.~S.~Krauser, U.~Ebling, N.~Fl\"aschner, J.~Heinze, K.~Sengstock, M.~Lewenstein, A.~Eckardt, and C.~Becker,
\textit{Giant spin oscillations in an ultracold Fermi sea},
Science \textbf{343}, 157 (2014)

\bibitem{Gallis1990}
M.~R.~Gallis, and G.~N.~Fleming,
\textit{Environmental and spontaneous localization},
Phys. Rev. A \textbf{42}, 38 (1990)

\bibitem{Hornberger2003}
K.~Hornberger, and J.~E.~Sipe,
\textit{Collisional decoherence reexamined},
Phys. Rev. A \textbf{68}, 012105 (2003)

\bibitem{Corruccini1972}
L.~R.~Corruccini, D.~D.~Osheroff, D.~M.~Lee, and R.~C.~Richardson,
\textit{Spin-wave phenomena in liquid \textsuperscript{3}He systems},
J. Low Temp. Phys. \textbf{8}, 229 (1972)

\bibitem{Johnson1984}
B.~R.~Johnson, J.~S.~Denker, N.~Bigelow, L.~P.~L\'evy, J.~H.~Freed, and D.~M.~Lee,
\textit{Observation of Nuclear Spin Waves in Spin-Polarized Atomic Hydrogen Gas},
Phys. Rev. Lett. \textbf{52}, 1508 (1984)

\bibitem{Bashkin1981}
E.~P.~Bashkin,
\textit{Spin waves in polarized paramagnetic gases},
JETP Lett. \textbf{33}, 8 (1981)

\bibitem{Levy1984}
L.~P.~L\'evy, and A.~E.~Ruckenstein
\textit{Collective Spin Oscillations in Spin-Polarized Gases: Spin-Polarized Hydrogen},
Phys. Rev. Lett. \textbf{52}, 1512 (1984)

\bibitem{Lhuillier1982I}
C.~Lhuillier, and F.~Lalo\"e,
\textit{Transport properties in a spin polarized gas, I},
J. Phys. (Paris) \textbf{43}, 197 (1982)

\bibitem{Lhuillier1982II}
C.~Lhuillier, and F.~Lalo\"e,
\textit{Transport properties in a spin polarized gas, II},
J. Phys. (Paris) \textbf{43}, 225 (1982)

\bibitem{OwersBradley1997}
J.~R.~Owers-Bradley,
\textit{Spin-polarized \textsuperscript{3}He-\textsuperscript{4}He liquids},
Rep. Prog. Phys. \textbf{60}, 1173 (1997)

\bibitem{Fuchs2003}
J.~N.~Fuchs, D.~M.~Gangardt, and F.~Lalo\"e,
\textit{Large amplitude spin waves in ultra-cold gases}, 
Eur. Phys. J. D \textbf{25}, 57 (2003)

\bibitem{Piechon2009}
F.~Pi\'echon, J.~N.~Fuchs, and F.~Lalo\"e,
\textit{Cumulative Identical Spin Rotation Effects in Collisionless Trapped Atomic Gases},
Phys. Rev. Lett. \textbf{102}, 215301 (2009)

\bibitem{Ebling2011}
U.~Ebling, A.~Eckardt, and M.~Lewenstein,
\textit{Spin segregation via dynamically induced long-range interactions in a system of ultracold fermions},
Phys. Rev. A \textbf{84}, 063607 (2011)

\bibitem{Fertig2005}
C. D.~Fertig, K. M.~O'Hara, J. H.~Huckans, S. L.~Rolston, W. D.~Phillips, and J. V.~Porto,
\textit{Strongly Inhibited Transport of a Degenerate 1D Bose Gas in a Lattice},
Phys. Rev. Lett. \textbf{94}, 120403 (2005)

\bibitem{Timmermans2001}
E.~Timmermans,
\textit{Degenerate Fermion Gas Heating by Hole Creation},
Phys. Rev. Lett. \textbf{87}, 240403 (2001)

\bibitem{Natu2010}
S.~S.~Natu, and E.~J.~Mueller,
\textit{Spin waves in a spin-1 Bose gas},
Phys. Rev. A \textbf{81}, 053617 (2010)

\bibitem{Kavoulakis1998}
G. M.~Kavoulakis, C. J.~Pethick, and H.~Smith,
\textit{Relaxation Processes in Clouds of Trapped Bosons above the Bose-Einstein Condensation Temperature},
Phys. Rev. Lett. \textbf{81}, 4036-4039 (1998)

\bibitem{Massignan2005}
P.~Massignan, G. M.~Bruun, and H.~Smith,
\textit{Viscous relaxation and collective oscillations in a trapped Fermi gas near the unitarity limit},
Phys. Rev. A \textbf{71}, 033607 (2005)

\bibitem{BreitRabi}
G.~Breit and I.~I.~Rabi,
\textit{Measurement of Nuclear Spin},
Phys. Rev.  \textbf{38(11)}, 2082-2083 (1931)
 
\bibitem{Arimondo1977}
E.~Arimondo, M.~Inguscio, and P.~Violino,
\textit{Experimental determinations of the hyperfine structure in the alkali atoms},
Rev. Mod. Phys. \textbf{49} 31-75 (1977)

\bibitem{Olshanii1998} 
M.~Olshanii, 
\textit{Atomic Scattering in the Presence of an External Confinement and a Gas of Impenetrable Bosons},
Phys. Rev. Lett. \textbf{81}, 938 (1998)

\end{thebibliography}
\end{document}